\documentclass[prd,superscriptaddress,preprintnumbers,tightenlines,nofootinbib, eqsecnum,twocolumn]{revtex4-2}

\usepackage{amsmath}
\usepackage{amsfonts}
\usepackage{amssymb}
\usepackage{physics}
\usepackage{bm}
\usepackage{mathrsfs}
\usepackage{graphicx}
\usepackage[colorlinks]{hyperref}
\usepackage[usenames]{color}
\usepackage{mathtools}
\DeclarePairedDelimiter{\ceil}{\lceil}{\rceil}
\definecolor{darkgreen}{rgb}{0,0.5,0}
\hypersetup{urlcolor=darkgreen}
\usepackage{empheq}
\usepackage{ulem}
\hypersetup{
 unicode=false,          
 pdftoolbar=true,        
 pdfmenubar=true,        
 pdffitwindow=false,     
 pdfstartview={FitH},    
 pdftitle={My title},    
 pdfauthor={Author},     
 pdfsubject={Subject},   
 pdfcreator={Creator},   
 pdfproducer={Producer}, 
 pdfkeywords={keyword1} {key2} {key3}, 
 pdfnewwindow=true,      
 colorlinks=true,       
 linkcolor=red,          
 citecolor=cyan,        
 filecolor=magenta,      
 urlcolor=darkgreen,           
 linktocpage=true
}

\allowdisplaybreaks

\DeclareSymbolFontAlphabet{\mathrsfs}{rsfs}
\DeclareMathAlphabet{\mathcal}{OMS}{cmsy}{m}{n}

\newcommand{\be}{\begin{equation}}
\newcommand{\ee}{\end{equation}}
\newcommand{\bse}{\begin{subequations}}
\newcommand{\ese}{\end{subequations}}
\newcommand{\ba}{\begin{align}}
\newcommand{\ea}{\end{align}}
\newcommand{\nn}{\nonumber}
\newcommand{\p}{\partial}
\newcommand{\calO}{\mathcal{O}}
\newcommand{\di}{\mathrm{i}} 
\newcommand{\de}{\mathrm{e}}

\newcommand{\dI}{\mathrm{I}}
\newcommand{\dJ}{\mathrm{J}}

\newcommand{\dm}{\mathrm{m}}

\makeatletter
\g@addto@macro\bfseries{\boldmath}
\makeatother

\defcitealias{Bernard:2022noq}{paper~I}
\begin{document}

\nocite{Bernard:2022noq}
\title{Quasi-Keplerian parametrization for eccentric compact binaries in scalar-tensor theories at second post-Newtonian order and applications}

\author{David \textsc{Trestini}}\email{trestini@fzu.cz}
\affiliation{CEICO, Institute of Physics of the Czech Academy of Sciences, Na Slovance 2, 182 21 Praha 8, Czechia}

\date{\today}

\begin{abstract}
The generalized quasi-Keplerian parametrization for compact binaries on eccentric bound orbits is established at second post-Newtonian (2PN) order in a class of massless scalar-tensor theories. This result is used to compute the orbit-averaged flux of energy and angular momentum at Newtonian order, which means relative 1PN order beyond the leading-order dipolar radiation of scalar-tensor theories. The secular evolution of the orbital elements is then computed at 1PN order. At leading order, the closed form ``Peters and Mathews'' relation between the semimajor axis $a$ and the eccentricity $e$ is found to be independent of any scalar-tensor parameter, and is given by \mbox{$a \propto e^{4/3}/(1-e^2)$.} Finally, the waveform is obtained at Newtonian order in the form of a spherical harmonic mode decomposition, extending to eccentric orbits the results obtained \mbox{in \href{https://doi.org/10.1088/1475-7516/2022/08/008}{[JCAP 08 (2022) 008]}.}
\end{abstract}


\maketitle
\section{Introduction}

The LIGO-Virgo-KAGRA collaboration has so far detected more than 90 gravitational wave (GW) events~\cite{ligoscientific:2021djp}, and future ground-based and spaceborne GW detectors such as LISA, Einstein Telescope, and Cosmic Explorer open new horizons for the observation of the Universe. So far, all events are compatible with general relativity (GR), but current detectors only probe a small subset of binary systems, namely stellar-mass binaries near merger. Moreover, since events are detected by matched filtering under the premise that the waveform should be described by GR, it is not excluded that a subpopulation of gravitational wave events exhibiting large deviations from GR have been missed in the data due to selection biases~\cite{Magee:2023muf}. One approach to modeling alternative theories is to take a theory-agnostic, parametrized approach~\cite{Will:2018bme,Yunes:2009ke,BSat94,Mishra:2010tp}. In this paper, I instead opt for a theory-dependent approach, and derive predictions for the gravitational waveform within the class of massless scalar-tensor (ST) theories of gravity, which were first introduced by Jordan~\cite{jordan1955schwerkraft}, Fierz~\cite{Fierz:1956zz}, and Brans and Dicke~\cite{BransDicke}, and later generalized in~\cite{Nordtvedt:1970uv,Wagoner:1970vr}. Binary pulsar observations have already put strong constraints on the parameters entering the models~\cite{Dgef92,Fujii:2005,DeFelice:2010aj,DoublePulsar,Kramer:2021jcw}, but nonperturbative phenomena appearing in the later stages of the inspiral could potentially help evade these constraints, such as dynamical scalarization~\cite{Barausse:2012da,Palenzuela:2013hsa}. Due to the no-hair theorem in ST theory, valid for stationary \emph{isolated} black~holes~\cite{Hawking:1972qk,Sotiriou:2011dz}, one expects that the motion and radiation of \emph{binary} black holes~(BHs) are indistinguishable from those of the GR solution. To some extent, this expectation has been confirmed by numerical relativity calculations~\cite{Healy:2011ef}. Consequently, relevant theory-dependent tests for ST theory involve binary neutron stars (or more exotic extended compact objects) as well as asymmetrical BH-neutron star binaries~\cite{Niu:2021nic,Takeda:2023wqn,Tan:2023pkd}.

I will work within the post-Newtonian (PN) formalism, i.e., an approximation for weak gravitational fields and small orbital velocities, and I will consequently focus on the inspiraling phase of coalescing binary systems. Previous results in massless ST~theories using the PN expansion were derived using (i) effective field theory methods~\cite{Dgef92,Dgef93,Dgef94,Dgef96,Dgef98}, (ii)  the direct integration of the relaxed field equations (DIRE) method~\cite{Wiseman:1992dv,WW96,Mirshekari:2013vb,Lang:2013fna,Lang:2014osa}, and (iii) the post-Newtonian, multipolar post-Minkowskian (PN-MPM) scheme~\cite{Bernard:2018hta,Bernard:2018ivi,Bernard:2022noq,Bernard:2019yfz,Bernard:2023eul}. It is the latter scheme that I will use hereafter, as it has been successfully used in GR to compute the energy flux and GW phasing at very high fourth-and-a-half post-Newtonian (4.5PN) order~\cite{bfhlt_letter,bfhlt_ii,TB23_ToM,TLB22,MBF16,MQ4PN_jauge,Henry:2021cek,MQ4PN_renorm,MQ4PN_IR,MHLMFB20}. 

Gravitational radiation in ST theories was first studied by~\cite{Dgef92,Dgef93,Damour:1996ke,Dgef96,Dgef98} in an alternative (but equivalent) formulation of the action. The equations of motion and conserved energy and angular momentum were first obtained at 2.5PN order in~\cite{Mirshekari:2013vb}, then at 3PN order in~\cite{Bernard:2018hta,Bernard:2018ivi}, including the contribution of the (dipolar) tail term. The waveform and flux was first obtained in the tensor sector at 2PN~\cite{Lang:2013fna} and in the scalar sector at 1PN~\cite{Lang:2014osa} beyond quadrupolar radiation (i.e. relative 2PN order beyond the leading-order scalar dipolar radiation). The flux was then reduced to circular orbits at 2PN in tensor sector and 1PN in the scalar sector~\cite{Sennett:2016klh}. The phase at 1PN and the tensor spherical harmonic modes at 2PN were computed as well~\cite{Sennett:2016klh}. Finally, the complete (tensor and scalar) flux for generic orbits was computed at 1.5PN in~\cite{Bernard:2022noq}, along with its reduction to circular orbits. Moreover, the scalar and tensor spherical harmonic modes were computed for circular orbits at 1PN order, and the phase at 1.5PN order~\cite{Bernard:2022noq}. The equations of motion and gravitational radiation has also been studied in extensions to massless scalar-tensor theories, e.g.~massive ST theory~\cite{Diedrichs:2023foj}, massless multiscalar-tensor theory~\cite{Schon:2021pcv}, Einstein-\ae ther theory \cite{Taherasghari:2023rwn}, scalar-Gauss-Bonnet theory~\cite{Shiralilou:2021mfl}, Horndeski theory~\cite{Higashino:2022izi,Wu:2023jwd}, etc.

In GR, gravitational waves generated by eccentric binaries were first studied at Newtonian order in the seminal work of Peters and Mathews~\cite{PM63,Peters64}. An elegant generalization of the Kepler solution to first post-Newtonian (1PN) order, dubbed the quasi-Keplerian (QK) parametrization, was then introduced by Damour and Deruelle~\cite{DD85}. This parametrization was then generalized for spinless systems at 2PN~\cite{DS88,SW93}, 3PN~\cite{MGS04} and finally 4PN order~\cite{ChoTanay2022}, including a semianalytical treatment of the tail term (although certain zero-average, oscillatory terms are ignored). The gravitational waveform and energy flux were first computed at 1PN order in~\cite{WagW76,BS89} and the 1PN angular momentum flux, waveform and secular evolution of the orbital parameters were computed in~\cite{JS92}. The hereditary tail effects entering at 1.5PN were first dealt with in the case of eccentric orbits in~\cite{BS93,RS97}. At 2PN order, the  energy and angular momentum fluxes, the secular evolution of the orbital elements and the waveform were computed by~\cite{Gopakumar:1997bs}. At 3PN, these same quantities were computed in the series of works~\cite{ABIQ08tail,ABIQ08,ABIS09}, which also treat higher-order tails and the memory integral that enters the angular momentum flux at this order. Moreover, the amplitudes of the spherical harmonic modes were computed in~\cite{Mishra:2015bqa, Boetzel:2019nfw, Ebersold:2019kdc}.  Partial results at 4PN for the waveform and orbital element evolution are given in~\cite{ChoTanay2022}.

Fewer results have been derived for eccentric systems in the context of ST theories. The parametrized post-Keplerian (PKK) parametrization was introduced in~\cite{Damour:1991rd}, and applied at leading order to obtain some predictions for tensor-multiscalar theories~\cite{Damour:1996ke}. In the case of monoscalar-tensor theories, the effects have been briefly studied in~\cite{Cardoso:2020iji}, as well as in~\cite{Khalil:2022rlq,Khalil:2022sii}. In the case of ST theories with screening generated by a potential, eccentric orbits have been investigated more in detail in~\cite{Zhang:2018prg}.

This work is a direct continuation of Ref.~\cite{Bernard:2022noq}, hereafter \citetalias{Bernard:2022noq}, and aims at systematically extending the results for circular orbits therein to the case of an eccentric orbit with the help of the QK parametrization. The system is assumed to be dipole driven, and the quadrupole-driven case is left for further investigation (see~\cite{Sennett:2016klh} for a discussion). After a brief notational subsection, the main features of the ST theories under consideration are introduced in Section~\ref{sec:STtheory}, as well as the MPM construction. Section~\ref{sec:PK} is devoted to the derivation of the QK parametrization at 2PN order in the context of alternative theories of gravity. Generic results are first derived in a class of alternative theories that I will define precisely.  Then, these results are specialized to the case of ST theories. The reason for this split is to provide generic expressions that will, in the future, facilitate the computation of the QK parametrization for other theories of gravity.  In Section~\ref{sec:flux}, the orbit-averaged fluxes of energy and angular momentum are computed at Newtonian order, which corresponds to the next-to-leading order beyond the dominant dipolar radiation. Section~\ref{sec:orbital_elements} is devoted to the computation of the evolution of the orbital elements. I first focus on the $-1$PN, leading order terms, for which there is no ambiguity in the definition of the semimajor axis $a$ and eccentricity $e$. In this case, I find closed form expressions akin to those of Peters and Mathews~\cite{PM63,Peters64}, such as $a\propto e^{4/3}/(1-e^2)$, and compare the orbital decay between ST theory and GR in the leading-order case. Then, I compute the 1PN expression of the two ordinary differential equations (ODEs) describing the time evolution of the pair of variables $(x,e_t)$. Since all other orbital parameters are explicitly given in this paper in terms of this pair of variables, their time evolution is straightforward to obtain. Finally, Section~\ref{sec:waveform} describes the waveform at (next-to-leading) Newtonian order as a mode decomposition over spherical harmonics, which is a useful result for numerical relativists.

\subsection*{Main notations and summary of parameters}
\label{sec:not}

The convention in this work is that all stated PN orders are, by default, relative to the the Newtonian dynamics and the standard quadrupole radiation in GR. Thus, the dominant dipole radiation enters at~$-0.5$PN order in the waveform and at~$-1$PN order in the energy flux. The the QK parametrization at second post-Newtonian order (``second post-Keplerian order'' or 2PK for short) computed in the paper is therefore next-to-next-to-leading-order, while the Newtonian fluxes and waveforms are next-to-leading order.

After a 3+1 decomposition of spacetime, and using the convention that boldface letters represent three-dimensional Euclidean vectors, the field-point spatial vector in the center-of-mass (CM) frame is denoted by $\bm{X} = R \,\bm{N}$, where $\bm{N}$ has unit norm. In spherical coordinates, the coordinates of the field point are denoted by $(R,\Theta,\Phi)$. Time is denoted by $t$, and retarded time is defined as  \mbox{$U=t-R/c$}, so as to not confuse it with the eccentric anomaly of the QK motion, $u$. The usual spherical harmonics $Y^{\ell \dm}(\Theta,\Phi)$ will be used, alongside the spin-weighted ones (with weight $-2$), $Y_{-2}^{\ell \dm}(\Theta,\Phi)$, where the integer $\dm$ should not be confused with the total mass $m$.

The positions of particles 1 and 2 are denoted by $\bm{y}_1$ and $\bm{y}_2$, and \mbox{$\bm{x} = \bm{y}_1-\bm{y}_2$} is the separation vector. The orbital radius is given by \mbox{$r = |\bm{x}|$}, and the unit vector \mbox{$\bm{n} = \bm{x}/r$} is introduced.  The relative velocity is given by \mbox{$\bm{v}=\dd\bm{x}/\dd t$}. For a nonprecessing eccentric compact binary in the CM frame, one can then introduce the orthonormal triad \mbox{$(\bm{n}, \bm{\lambda},\bm{\ell})$} such that $\bm{\lambda}$ lies in the orbital plane, \mbox{$\bm{\lambda}\cdot\bm{v}>0$} and \mbox{$\bm{\ell} = \bm{n}\times\bm{\lambda}$}. Introducing some reference, nonrotating reference triad \mbox{$(\bm{n}_0, \bm{\lambda}_0,\bm{\ell})$}, one can then describe the orbital motion by the polar coordinates\footnote{The notation $\phi$ is used both for the scalar field and the polar angle, but its meaning will always be clear in context.} $(r,\phi)$ such that $\bm{n} = \cos \phi \,\bm{n}_0+\sin \phi \,\bm{\lambda}_0$. Notice that the relative velocity is given by \mbox{$\bm{v} = \dot{r} \bm{n} + r\dot{\phi} \bm{\lambda}$}. The following relations then hold: \mbox{$\bm{n}\cdot\bm{v}=\dot{r}$}, \mbox{$v^2=\dot{r}^2 + r^2\dot{\phi}^2$}, and \mbox{$\bm{n}\times\bm{v}=r\dot{\phi}\bm{\ell}$}. 

The notation $L=i_1\cdots i_\ell$ stands for multi-index with $\ell$ spatial indices (and $K$ would stand for a multi-index with $k$ spatial indices). One can then write $\partial_L = \partial_{i_1}\cdots\partial_{i_\ell}$, $\partial_{aL-1} = \partial_a\partial_{i_1}\cdots\partial_{i_{\ell-1}}$ and so on;  similarly,  $n_{L} = n_{i_1}\cdots n_{i_\ell}$, $n_{aL-1} = n_a n_{i_1}\cdots n_{i_{\ell-1}}$. The symmetric trace-free (STF) part is indicated using a hat or angled brackets: for instance, $\mathrm{STF}_L [\partial_L] = \hat{\partial}_L = \partial_{\langle i_1} \partial_{i_2}\cdots \partial_{ i_\ell \rangle}$, $\mathrm{STF}_L [n_L] = \hat{n}_L = n_{\langle i_1} n_{i_2}\cdots n_{ i_\ell \rangle}$, and $\mathrm{STF}_L [x_L] = \hat{x}_L = r^\ell n_{\langle i_1} n_{i_2}\cdots n_{ i_\ell \rangle}$.
The $n$th time derivative of a function $F(t)$ is denoted $F^{(n)}(t) = \dd^n F / \dd t^n$.

The constant asymptotic value of the scalar field at spatial infinity is denoted $\phi_0$, and the normalized scalar field is defined by $\varphi\equiv\phi/\phi_{0}$.  The Brans-Dicke-like scalar function $\omega(\phi)$ is expanded around the asymptotic value $\omega_0 \equiv \omega(\phi_0)$ and  the mass functions $m_A(\phi)$ (see Section~\ref{subsec:fieldEquations} for a definition) are expanded around the asymptotic values $m_A \equiv m_A(\phi_0)$, where $A\in\{1,2\}$.  In the CM frame, one then defines the asymptotic total mass $m=m_1+m_2$, reduced mass $\mu = m_1 m_2 / m$, symmetric mass ratio $\nu= \mu/ m \in \ ]0,1/4]$, and relative mass difference $\delta = (m_1-m_2)/m \in \ [0,1[ $. Note that the asymptotic symmetric mass ratio and the relative mass difference are linked by the relation $\delta^2 = 1-4\nu$.

Following~\cite{Bernard:2018hta, Bernard:2022noq}, a number of parameters describing these expansions are introduced in~Table~\ref{table}. The ST parameters are defined directly from the expansions of the Brans-Dicke-like scalar function $\omega(\phi)$ and of the mass functions $m_A(\phi)$. The PN parameters are combinations of the ST parameters that naturally extend and generalize the usual PPN parameters to the case of a general ST theory~\cite{will1972conservation,Will:2018bme}. 
\begin{widetext}
\begin{small}
\begin{center}\begin{table}[h]
\begin{tabular}{|c||cc|}
	\hline
	& \multicolumn{2}{|c|}{\textbf{ST parameters}} \\[2pt]
	\hline &&\\[-10pt]
	general & \multicolumn{2}{c|}{$\omega_0=\omega(\phi_0),\qquad\omega_0'=\eval{\frac{\dd\omega}{\dd\phi}}_{\phi=\phi_0}, \qquad\omega_0''=\eval{\frac{\dd^2\omega}{\dd\phi^2}}_{\phi=\phi_0},\qquad\varphi = \frac{\phi}{\phi_{0}},\qquad\tilde{g}_{\mu\nu}=\varphi\,g_{\mu\nu},$} \\[12pt]
	& \multicolumn{2}{|c|}{$\tilde{G} = \frac{G(4+2\omega_{0})}{\phi_{0}(3+2\omega_{0})},\qquad \zeta = \frac{1}{4+2\omega_{0}},$} \\[8pt]
	& \multicolumn{2}{|c|}{$\lambda_{1} = \frac{\zeta^{2}}{(1-\zeta)}\left.\frac{\dd\omega}{\dd\varphi}\right\vert_{\varphi=1},\qquad \lambda_{2} = \frac{\zeta^{3}}{(1-\zeta)}\left.\frac{\dd^{2}\omega}{\dd\varphi^{2}}\right\vert_{\varphi=1}, \qquad \lambda_{3} = \frac{\zeta^{4}}{(1-\zeta)}\left.\frac{\dd^{3}\omega}{\dd\varphi^{3}}\right\vert_{\varphi=1}.$} \\[7pt]
	\hline &&\\[-7pt]
	~sensitivities~ & \multicolumn{2}{|c|}{$s_A = \eval{\frac{\dd \ln{m_A(\phi)}}{\dd\ln{\phi}}}_{\phi=\phi_0},\qquad s_A^{(k)} = \eval{\frac{\dd^{k+1}\ln{m_A(\phi)}}{\dd(\ln{\phi})^{k+1}}}_{\phi=\phi_0},\qquad(A=1,2)$} \\[9pt]
    & \multicolumn{2}{|c|}{$s'_A = s_A^{(1)},\qquad s''_A = s_A^{(2)},\qquad s'''_A = s_A^{(3)},$} \\[5pt]
	& \multicolumn{2}{|c|}{$\mathcal{S}_+ = \frac{1-s_1 - s_2}{\sqrt{\alpha}}\,,\qquad \mathcal{S}_- = \frac{s_2 - s_1}{\sqrt{\alpha}}.$} \\[7pt]	\hline\hline 
	Order & \multicolumn{2}{|c|}{\textbf{PN parameters}} \\[2pt]
	\hline &&\\[-10pt]
	N & \multicolumn{2}{|c|}{$\alpha= 1-\zeta+\zeta\left(1-2s_{1}\right)\left(1-2s_{2}\right)$}   \\[5pt]
	\hline &&\\[-10pt]
	1PN & $\overline{\gamma} = -\frac{2\zeta}{\alpha}\left(1-2s_{1}\right)\left(1-2s_{2}\right),$ & Degeneracy \\[5pt]
	&~~$\overline{\beta}_{1} = \frac{\zeta}{\alpha^{2}}\left(1-2s_{2}\right)^{2}\left(\lambda_{1}\left(1-2s_{1}\right)+2\zeta s'_{1}\right),$~~~~&  $\alpha(2+\overline{\gamma})=2(1-\zeta)$ \\[5pt]
	& $\overline{\beta}_{2} = \frac{\zeta}{\alpha^{2}}\left(1-2s_{1}\right)^{2}\left(\lambda_{1}\left(1-2s_{2}\right)+2\zeta s'_{2}\right),$~~~~& \\[5pt]
	&  $\overline{\beta}_+ = \frac{\overline{\beta}_1+\overline{\beta}_2}{2}, \qquad \overline{\beta}_- = \frac{\overline{\beta}_1-\overline{\beta}_2}{2}.$ &  \\[5pt]
	\hline &\\[-10pt]
	2PN & $\overline{\delta}_{1} = \frac{\zeta\left(1-\zeta\right)}{\alpha^{2}}\left(1-2s_{1}\right)^{2}\,,\qquad \overline{\delta}_{2} = \frac{\zeta\left(1-\zeta\right)}{\alpha^{2}}\left(1-2s_{2}\right)^{2},$ & Degeneracy \\[5pt]
	&  $\overline{\delta}_+ = \frac{\overline{\delta}_1+\overline{\delta}_2}{2}, \qquad \overline{\delta}_- = \frac{\overline{\delta}_1-\overline{\delta}_2}{2},$ &  $16\overline{\delta}_{1}\overline{\delta}_{2} = \overline{\gamma}^{2}(2+\overline{\gamma})^{2}$\\[5pt]
	& $~~\overline{\chi}_{1} = \frac{\zeta}{\alpha^{3}}\left(1-2s_{2}\right)^{3}\left[\left(\lambda_{2}-4\lambda_{1}^{2}+\zeta\lambda_{1}\right)\left(1-2s_{1}\right)-6\zeta\lambda_{1}s'_{1}+2\zeta^{2}s''_{1}\right],~~$  &  \\[5pt]
	& $\overline{\chi}_{2} = \frac{\zeta}{\alpha^{3}}\left(1-2s_{1}\right)^{3}\left[\left(\lambda_{2}-4\lambda_{1}^{2}+\zeta\lambda_{1}\right)\left(1-2s_{2}\right)-6\zeta\lambda_{1}s'_{2}+2\zeta^{2}s''_{2}\right],$ &  \\[5pt]
	&  $\overline{\chi}_+ = \frac{\overline{\chi}_1+\overline{\chi}_2}{2}, \qquad \overline{\chi}_- = \frac{\overline{\chi}_1-\overline{\chi}_2}{2}.$ &  \\[5pt]
\hline
\end{tabular}
\caption{Summary of parameters for the general ST theory and notations for PN parameters. \label{table}}\end{table}
\end{center}
\end{small}\newpage
\end{widetext}


\section{Massless scalar-tensor theories}\label{sec:STtheory}


\label{subsec:fieldEquations}
As in \citetalias{Bernard:2022noq}, consider a generic class of ST theories in which a single massless scalar field~$\phi$ minimally couples to the metric~$g_{\mu\nu}$. It is described by the action
\begin{align}\label{eq:STactionJF}
S_{\mathrm{ST}} &= \frac{c^{3}}{16\pi G} \int\dd^{4}x\,\sqrt{-g}\left[\phi R - \frac{\omega(\phi)}{\phi}g^{\alpha\beta}\p_{\alpha}\phi\p_{\beta}\phi\right] \nn\\*
&\qquad\quad+S_{\mathrm{m}}\left(\mathfrak{m},g_{\alpha\beta}\right)\,,
\end{align}
where $R$ and $g$ are, respectively, the Ricci scalar and the determinant of the metric, $\omega$ is a function of the scalar field and $\mathfrak{m}$ stands generically for the matter fields. The action for the matter $S_{\mathrm{m}}$ is a function only of the matter fields and the metric. A major difference in ST theories compared to GR is that, as a consequence of the breaking of the strong equivalence principle, one has to take into account the internal gravity of each body. Indeed, the scalar field determines the effective gravitational constant, which in turn affects the competition between gravitational and nongravitational forces within the body. Thus, the value of the scalar field has an indirect influence on the size of the compact body and on its internal gravity. Here, I follow the approach pioneered by Eardley~\cite{Eardley1975} (see also~\cite{Nordtvedt:1990zz}) and take for $S_{\mathrm{m}}$ the effective action for $N$ nonspinning point particles with the masses $m_A(\phi)$ depending in an unspecified manner on the value of the scalar field at the location of the particles, i.e.,
\be\label{eq:matteract}
S_{\mathrm{m}} = - c \sum_{A} \int\,m_{A}(\phi) \sqrt{-\left(g_{\alpha\beta}\right)_{A}\dd y_{A}^{\alpha}\,\dd y_{A}^{\beta}}\,,
\ee
where $y_A^\alpha$ denote the space-time positions of the particles, and $\left(g_{\alpha\beta}\right)_{A}$ is the metric evaluated\footnote{Divergences are treated with Hadamard regularization~\cite{BFreg}, which is equivalent, at this order, to dimensional regularization~\cite{BDE04,BDEI05dr}.} at the position of particle~$A$. Thus, the matter action depends indirectly on the scalar field, and the sensitivities of the particles to variations in the scalar field are defined by 
\be\label{sAk}
s_A^{(k)} \equiv \eval{\frac{\dd^{k+1}\ln{m_A(\phi)}}{\dd(\ln{\phi})^{k+1}}}_{\phi=\phi_0}\,, 
\ee
where $s_A \equiv  s_A^{(1)}$ and $\phi_0$ are the values of the scalar field at spatial infinity that is assumed to be constant in time, i.e., the cosmological evolution is neglected. For stationary black holes, since all information regarding the matter which formed the black hole has disappeared behind the horizon, the mass can depend only on the Planck scale, $m_A \propto M_\text{Planck} \propto G^{-1/2} \propto \phi^{1/2}$ hence $s^\text{BH}_A={1}/{2}$.
The action~\eqref{eq:STactionJF} is usually called the Jordan-frame action, as the matter only couples to the Jordan or ``physical'' metric $g_{\alpha\beta}$. It is then very useful to introduce a rescaled scalar field and a conformally related metric,
\be\label{eq:def_gt}
\varphi\equiv \frac{\phi}{\phi_{0}}\qquad\mathrm{and}\qquad\tilde{g}_{\alpha\beta}\equiv \varphi\,g_{\alpha\beta}\,,
\ee
such that the physical and conformal metrics have the same asymptotic behavior at spatial infinity. Quantities expressed in terms of the pair $(\varphi, \tilde{g}_{\alpha\beta})$ are said to be in the ``Einstein frame.''
%
%
From these, one defines the scalar and metric perturbation variables  $\psi\equiv\varphi-1$ and $h^{\mu\nu}\equiv \sqrt{-\tilde{g}}\tilde{g}^{\mu\nu}-\eta^{\mu\nu}$, where $\eta^{\mu\nu}~=~\text{diag}(-1,1,1,1)$ is the Minkowski metric. The field equations then read
\begin{subequations}\label{eq:rEFE}
\begin{align}
& \Box_{\eta}\,h^{\mu\nu} = \frac{16\pi G}{c^{4}}\tau^{\mu\nu}\,,\\
& \Box_{\eta}\,\psi = -\frac{8\pi G}{c^{4}}\tau_{s}\,,
\end{align}
\end{subequations}
where $\Box_{\eta}$ denotes the ordinary flat space-time d'Alembertian operator, and where the source terms read
\begin{subequations}
\begin{align}\label{eq:taumunu}
& \tau^{\mu\nu} = \frac{\varphi}{\phi_{0}} (- g) T^{\mu\nu} +\frac{c^{4}}{16\pi G}\Lambda^{\mu\nu}[h,\psi]\,,\\
\label{taus} & \tau_{s} = -\frac{\varphi}{\phi_{0}(3+2\omega)}\sqrt{-g}\left(T-2\varphi\frac{\p T}{\p \varphi}\right) -\frac{c^{4}}{8\pi G}\Lambda_s[h,\psi]\,.
\end{align}
\end{subequations}
%
Here $T^{\mu\nu}= 2 (-g)^{-1/2}\delta S_{\mathrm{m}}/\delta g_{\mu\nu}$ is the matter stress-energy tensor, $T\equiv g_{\mu\nu}T^{\mu\nu}$ is its trace and $\p T/\p \varphi$ is defined as the partial derivative of $T(g_{\mu\nu}, \varphi)$ with respect to $\varphi$ holding $g_{\mu\nu}$ constant. Moreover, $\Lambda^{\mu\nu}$ and $\Lambda^s$ are given explicitly in (2.8) and (2.9) of \citetalias{Bernard:2022noq} as functionals of the $h^{\mu\nu}$ and $\psi$ that are at least quadratic in the fields.  

The field equations of \eqref{eq:rEFE} are solved using the PN-MPM construction \cite{BD86,Blanchet:2013haa}. The reader is invited to refer to \citetalias{Bernard:2022noq} for a detailed description of this construction in the case of ST theories.  For the purposes of this work, I will only need to use the expressions of the so-called source moments $\dI_L$, $\dJ_L$  and $\dI_L^s$, which are pure functions of retarded time $U$ that are STF in their indices and uniquely parametrize the linearized metric  $h_1^{\mu\nu}$ and the linearized scalar field $\psi_1$. For example, the linearized scalar field reads explicitly :
\be 
\psi_1 = - \frac{2}{c^2}\sum_{\ell=0}^{+\infty}\frac{(-)^\ell}{\ell!}\,\partial_L\!\left[\frac{\dI_L^s(U)}{R}\right]\,.\label{psi1}
\ee
A feature of ST theory is that the scalar monopole $\dI^s$ is not constant but its time-variation will be a small PN effect, i.e., $\dd \dI^s/\dd t = \mathcal{O}(c^{-2})$. This is why the leading-order radiation is dipolar, and not monopolar. It is thus practical to define 
\bse\be
\dI^s(U) = \frac{1}{\phi_0}\left[ m^s+\frac{E^s(U)}{c^2} \right]\,,
\ee
where 
\be 
m^s = - \frac{1}{3+2\omega_0}\sum_A m_A (1- 2s_A)
\ee\ese
is constant and $E^s(U)$ is the time-varying PN correction.

The gravitational and scalar waves, when viewed by an asymptotic observer [i.e. $G m/(c^2 R_\mathrm{obs}) \ll 1$, where $R_\mathrm{obs}$ is the distance of the observer to the source], admits a multipolar decomposition as well. Introducing some radiative coordinates $(T,R,\bm{N})$ [or alternatively  $(T,R,\Theta,\Phi)$], which ensure that the asymptotic structure of the waveform is parametrized by three families of \emph{radiative} moments~\cite{Blanchet:2013haa, Bernard:2022noq}, which are denoted $\mathcal{U}_L$, $\mathcal{V}_L$ and $\mathcal{U}^s_L$. It reads in a transverse-traceless gauge
\begin{subequations}\label{eq:radiative_expansion}
\begin{align}
 	h_{ij}^\text{TT} &= - \frac{4G}{c^2 R} \perp^\text{TT}_{ijab} \sum_{\ell=2}^{+\infty} \frac{1}{c^\ell \ell!}\Bigg[ N_{L-2}\,\mathcal{U}_{ab L-2}(U) \nn\\*
 	&\qquad- \frac{2\ell}{c(\ell+1)} N_{c L-2} \epsilon_{cd(a}\mathcal{V}_{b)dL-2}(U)\Bigg] + \mathcal{O}\Big(\frac{1}{R^2}\Big)\,,\label{seq:tensor_radiative_expansion}\\
 	\psi &= - \frac{2G}{c^2 R}\sum_{\ell=0}^{+\infty} \frac{1}{c^\ell \ell!} N_L \mathcal{U}_L^s(U) + \mathcal{O}\Big(\frac{1}{R^2}\Big)\,, \label{seq:scalar_radiative_expansion}
\end{align}
\end{subequations}
where \mbox{$\perp_{ijab}^\mathrm{TT} = \perp_{a(i}\perp_{j)b}-\frac{1}{2}\!\perp_{ij}\perp_{ab}$} with \mbox{$\perp_{ij} =\delta_{ij}-N_i N_j$}.
The radiative moments are related to the source moments by the relations
\begin{align}
 \mathcal{U}_L(U) &=  \overset{(\ell)}{\dI}_{\!\!L}(U)+\mathcal{O}\left(\frac{1}{c^3}\right)\,,\\
 \mathcal{V}_L(U) &=  \overset{(\ell)}{\dJ}_{\!\!L}(U) +\mathcal{O}\left(\frac{1}{c^3}\right)\,,\\
\mathcal{U}^s_L(U) &= \overset{(\ell)}{\dI^s }_{\!\!L}(U)+\mathcal{O}\left(\frac{1}{c^3}\right)\,,
 \end{align}
where the small $\mathcal{O}(1/c^3)$ corrections correspond to the nonlinear corrections of the MPM construction, such as tails or memory integrals (see \citetalias{Bernard:2022noq} for details). In this work, only the linear part of the MPM construction is needed.

 \section{The quasi-Keplerian parametrization}	
 \label{sec:PK}
 Let us now turn to the computation of the QK~parametrization for eccentric orbits at second post-Newtonian order, which is the main result of this work. The goal is to obtain a closed form expression for the motion of bound orbits obeying the 2PN conservative equations of motion of scalar-tensor theory. The motion is parametrized only in terms of two free variables, e.g. the energy and angular momentum. In Section~\ref{subsec:PK_general}, I define the QK~parametrization  and obtain some generic results that depend only on the functional relation linking the energy and angular momentum of the binary system to the relative trajectories and velocities. These results are then specialized in Section~\ref{subsec:PK_ST} to the case of the scalar-tensor theory considered in this work, and yield the expression of the PK parameters (e.g., the semimajor axis $a_r$ and eccentricities $e_r$, $e_t$, and $e_\phi$) in terms of the energy and angular momentum of the binary. The reason for this split is that the generic results can be very useful to obtain the QK parametrization of many other alternative theories of gravity.
\subsection{General case}
 \label{subsec:PK_general}
Consider the conservative dynamics of spinless binary system, characterized by the expression of the conserved energy $E$ and angular momentum~$\bm{J}$ (whose norm is denoted~$J = |\bm{J}|$) in terms of the relative position $\mathbf{r}$ and velocity $\mathbf{v}$ of the two objects. If the motion is planar, it is in principle possible to invert these relation to obtain the dynamics in the following form:
\bse \label{eq:rdot2phidot} \begin{align}
\dot{r}^2 &= s^{-4}\dot{s}^2= \mathcal{R}(s, E, J)  \label{seq:rdot2}\,, \\
\dot{\phi} &= \mathcal{S}(s,E, J) \label{seq:phidot} \,,
\end{align}\ese
where $s \equiv 1/r$. In this section, I thus assume that we are working in a \emph{generic theory of gravity} that can be studied perturbatively using the PN methodology and whose 2PN equations of motion  in the CM frame can be cast in the form of \eqref{eq:rdot2phidot}, where $\mathcal{R}$ and $\mathcal{S}$ are fifth-order polynomials in $s$, which have the following structure (see Section~\ref{subsec:PK_ST} for how this structure arises in the particular case of ST theories):
\bse\label{eq:RS}\begin{align}
\mathcal{R}(s, E, J) &= A+ 2B s+C s^2+D_1 s^3 +D_2 s^4 +D_3 s^5 \label{seq:R}\,,\\
\mathcal{S}(s, E, J) &= F s^2+I_1 s^3 +I_2 s^4 +I_3 s^5 \label{seq:S}\,.
\end{align}\ese
Here,  $A$, $B$, $C$, and $F$ are constants of order $\calO(1)$, $D_1$ and $I_1$ are of order $\calO(c^{-2})$, and $D_2$,   $D_3$,  $I_2$, and $I_3$ are of order $\calO(c^{-4})$.  In particular, note the absence of any logarithmic term (which would typically appear in standard harmonic coordinates at higher order) or nonlocal tail terms.
In GR, the expressions of these coefficients are well known~\cite{DD85, DS88, SW93}, while their expressions in  massless ST theories will be computed in Section~\ref{subsec:PK_ST}. Thanks to the results given in~\eqref{eq:PK_parameters_general} of this Section, it will suffice to obtain an equation of the form \eqref{eq:RS} for any given theory of gravity to automatically know the QK parametrization at 2PN order.

Dropping from now on the dependencies in $E$ and $J$, the previous assumptions  imply that the polynomial $\mathcal{R}$ has exactly two roots that are nonzero in the $c \rightarrow \infty$ limit: denote these by $s_p = 1/r_p$ and $s_a =1/r_a$, where ``$p$'' and ``$a$'' stand for periastron and apastron, respectively (thus $r_p<r_a$ and $s_p>s_a$). One then obtains the factorization
\be \mathcal{R}(s) = (s-s_a)(s_p-s)\widetilde{\mathcal{R}}(s) \,, \label{eq:R_factorized} \ee
where $\widetilde{\mathcal{R}}(s)$ is a third-order polynomial. One can iteratively compute these roots to desired 2PN precision. 

In the QK motion, the orbital period $P$ is defined as the time of return to the periastron, which reads
\bse \label{eq:P_Phi_integrals} \be P = 2 \int_{s_a}^{s_p} \frac{\dd s}{s^2\sqrt{\mathcal{R}(s)}} \,. \label{seq:P_integral} \ee
From this it is useful to define the mean motion $n = 2\pi/P$ and the mean anomaly $\ell = n(t-t_0)$. During one period $P$, the true anomaly $\phi$ does not increase by $2\pi$, but by a bit more, due to the well known advance of the periastron. The increase in the true anomaly per period $P$ is given by
\be \Phi = 2 \int_{s_a}^{s_p} \frac{\dd s\,\mathcal{S}(s)}{s^2\sqrt{\mathcal{R}(s)}} \,. \label{seq:Phi_integral} \ee\ese
One defines from it the quantity $K = \Phi/(2\pi)$, as well as $k=K-1$ and $\Delta\Phi = 2\pi k$.

These periods are most efficiently computed to some finite PN accuracy using complex analysis methods, which are described in Appendix \ref{app:complex}. The end result (given in machine-readable form in the Supplementary Material~\cite{SuppMaterial}) at 2PN accuracy reads

\bse\label{eq:n_K_general}\begin{align}
n &= \frac{(-A)^{3/2}}{B} + \frac{A^3(12 B D_3 + 3 D_1^2-4 C D_2)}{8 B^2(-C)^{5/2}} \,, \\
K &=  \frac{F}{\sqrt{-C}} + \frac{B(3 F D_1-2 C I_1)}{2 (-C)^{5/2}}\nn\\*
&\quad + \frac{1}{16 (-C)^{9/2}}\Big(105 B^2 F D_1^2 - 15 A C F D_1^2  \nn\\*
& \qquad - 60 B^2 C F D_2 + 12 A  C^2 F D_2  + 140 B^3 F D_3  \nn\\*
& \qquad - 60 A B C F D_3 - 60  B^2 C D_1 I_1 + 12 A C^2  D_1 I_1 \nn\\*
& \qquad + 24 B^2 C^2  I_2  - 8 A C^3  I_2 - 40 B^3 C I_3 + 24 A B C^2 I_3  \Big) \,,
\end{align}\ese
and the expression for $n$ agrees at lowest order with (3.5a) of~\cite{DD85}.
The QK parametrization is formally valid as soon as one can write down Eq.~\eqref{eq:RS}. At 2PN order, it reads~\cite{DS88,SW93}
\bse \label{eq:PK_equations}
\begin{align}
r &= a_r(1- e_r \cos(u)) \,, \label{seq:radial_equation}\\
\ell=n(t-t_0)&= u - e_t \sin(u) + f_t \sin(v) + g_t(v-u) \,,  \label{seq:kepler_equation}\\
\frac{2\pi}{\Phi}(\phi-\phi_0) &= v + f_\phi \sin(2v) + g_\phi \sin(3v) \,, \label{seq:angular_equation}
\end{align}
\ese
where $u$ is the eccentric anomaly, and 
\begin{align}\label{eq:v}
v &= 2 \arctan\Bigg[ \sqrt{\frac{1+e_\phi}{1-e_\phi}}\tan\Big(\frac{u}{2}\Big) \Bigg] + 2\pi\bigg\lfloor \frac{u+\pi}{2\pi}\bigg\rfloor \nonumber\\
&= u + 2 \arctan\Bigg[\frac{\beta(e_\phi) \sin(u)}{1-\beta(e_\phi) \cos(u)}\Bigg] \,,
\end{align}
with $\beta(e) \equiv e/\left(1+\sqrt{1-e^2}\right)$. Note that the term featuring a floor function is constant by parts, always a multiple of~$2\pi$, and is introduced only to ensure continuity over the whole real axis.

The detailed construction of the different PK parameters is described in Appendix \ref{app:PKconstruction}. Here, let me simply highlight the following facts. First, $g_t$, $f_t$, $g_\phi$ and $f_\phi$ are all~$\sim \calO(c^{-4})$, and vanish at the 1PN level. Second, the three eccentricities $e_r$, $e_\phi$ and $e_t$ are equal at Newtonian level. Third, one has the straightforward relations \mbox{$a_r = (s_p+s_a)/(2s_a s_p)$} and \mbox{$e_r = (s_p-s_a)/(s_p+s_a)$}.

The expressions for the  values of the different PK parameters in terms of $A$, $B$, $C$, $D_n$, $F$, and $I_n$ are presented hereafter. The expressions in the case of the massless ST theories of \eqref{eq:STactionJF} are given in Appendix \ref{app:PKparameters}. Both results are provided in machine-readable form in the Supplementary Material~\cite{SuppMaterial}.

\begin{widetext}
\bse\label{eq:PK_parameters_general}\begin{align}
a_r &= - \frac{B}{A} + \frac{D_1}{2C} + \frac{2 B D_1^2 - 2 B C D_2 + 4 B^2 D_3 - A C D_3}{2 C^3} + \mathcal{O}\left(\frac{1}{c^4}\right) \,, \\
e_r &= \sqrt{1-\frac{AC}{B^2}}+\frac{A D_1 (2B^2-AC)}{2B^2 C \sqrt{B^2-AC}}\nn\\*
&\quad + \frac{A}{8 B^3 C^3(B^2 - A C)^{3/2}}\Big(16 B^6 D_1^2 - 24 A B^4 C D_1^2 + 5 A^2 B^2 C^2 D_1^2 + 2 A^3 C^3 D_1^2 - 16 B^6 C D_2 + 28 A B^4 C^2 D_2\nn\\*
& \qquad\qquad\qquad\qquad\qquad\qquad - 12 A^2 B^2 C^3 D_2 +32 B^7 D_3 - 64 A B^5 C D_3 + 36 A^2 B^3 C^2 D_3 - 4 A^3 B C^3 D_3   \Big) + \mathcal{O}\left(\frac{1}{c^4}\right) \,,\\
e_t &= \sqrt{1-\frac{AC}{B^2}}+ \frac{A D_1}{2 C \sqrt{B^2 - A C}} + \frac{(-A)^{3/2}\sqrt{B^2 - A C}}{8 B^2 (-C)^{5/2}}\left(4 C D_2 - 12 B D_3 -3D_1^2\right) \nn\\*
& \qquad +\frac{A}{8 B C^3 (B^2 - A C)^{3/2}}\Big(8 B^4 D_1^2 -12 A B^2 C D_1^2 + 3 A^2 C^2 D_1^2 -8 B^4 C D_2+ 12 A B^2 C^2 D_2  \nn\\*
& \qquad\qquad\qquad\qquad\qquad\qquad  - 4 A^2 C^3 D_2 + 16 B^5 D_3 -28 A B^3 C D_3 + 12 A^2 B C^2 D_3\Big)+ \mathcal{O}\left(\frac{1}{c^4}\right) \,,\\
e_\phi &= \sqrt{1-\frac{AC}{B^2}}+ \frac{A(3B^2 F D_1 - 2 A C F D_1 - 2 B^2 C I_1 + 2 A C^2 I_1)}{2B^2 C F \sqrt{B^2 - A C}} \nn\\*
&\quad  + \frac{A}{8 B^3 C^3 F^2 (B^2-AC)^{3/2}}\Big(26 B^6 F^2 D_1^2 - 36 A B^4 C F^2 D_1^2 + A^2 B^2 C^2 F^2 D_1^2 + 8 A^3 C^3 F^2 D_1^2 -32 B^6 C F^2 D_2 \nn\\*
& \qquad\qquad\qquad\qquad\qquad\qquad\quad+60 A B^4 C^2 F^2 D_2 - 28 A^2 B^2 C^3 F^2 D_2 + 79 B^7 F^2 D_3 - 165 A B^5 C F^2 D_3 \nn\\*
& \qquad\qquad\qquad\qquad\qquad\qquad\quad + 97 A^2 B^3 C^2 F^2 D_3 -11 A^3 B C^3 F^2 D_3 + 8 B^6 C F D_1 I_1 -36 A B^4 C^2 F D_1 I_1 \nn\\*
& \qquad\qquad\qquad\qquad\qquad\qquad\quad + 44 A^2 B^2 C^3 F D_1 I_1 - 16 A^3 C^4 F D_1 I_1 -16 B^6 C^2 I_1^2 + 40 A B^4 C^3 I_1^2 - 32 A^2 B^2 C^4 I_1^2  \nn\\*
& \qquad\qquad\qquad\qquad\qquad\qquad\quad + 8 A^3 C^5 I_1^2 +16 B^6 C^2 F I_2 - 32 A B^4 C^3 F I_2 + 16 A^2 B^2 C^4 F I_2 - 30 B^7 C F I_3  \nn\\*
& \qquad\qquad\qquad\qquad\qquad\qquad\quad+ 66 A B^5 C^2 F I_3 - 42 A^2 B^3 C^3 F I_3 + 6 A^3 B C^4 F I_3 \Big) + \mathcal{O}\left(\frac{1}{c^4}\right) \,,\\
f_t &= \frac{(-A)^{3/2} D_3 \sqrt{B^2 - AC}}{2 B (-C) ^{5/2}} + \mathcal{O}\left(\frac{1}{c^4}\right) \,, \\
f_\phi &= \frac{B^2 - AC}{32 C^4 F^2}\Big(F^2 D_1^2 - 4 C F^2 D_2 +20 B F^2 D_3 + 4 C F D_1 I_1 - 8 C^2  I_1^2 + 8 C^2  F I_2  - 24 B  C F I_3  \Big) + \mathcal{O}\left(\frac{1}{c^4}\right) \,,\\
g_t &=\frac{(-A)^{3/2}(3 D_1^2 -4 C D_2+12 B D_3 )}{8 B (-C) ^{5/2}} + \mathcal{O}\left(\frac{1}{c^4}\right)  \,, \\
g_\phi &=  \frac{(B^2 - AC)^{3/2}(F D_3-2 C I_3)}{24 C^4  F} + \mathcal{O}\left(\frac{1}{c^4}\right) \,.
\end{align}\ese
\end{widetext}

\subsection{The case of ST theory}
\label{subsec:PK_ST}
The results of the previous section are now specialized to the case of the ST theory described by the action \eqref{eq:STactionJF}. The starting point is to consider the expressions of the conserved energy $E(\mathbf{x},\mathbf{v})$ and angular momentum $J(\mathbf{x},\mathbf{v})$ in the CM frame for ST theories in terms of the relative positions and velocities of the two particles. The expressions of the energy and angular momentum were computed at 3PN order in (4.1) and (4.2) of~\cite{Bernard:2018ivi}, but only the 2PN expressions are needed here. Note that these relations are given in terms of the standard harmonic coordinates, and since they exhibit no pathological logarithms at this order, there is no need to introduce modified harmonic coordinates. The case of Arnowitt-Deser-Misner coordinates is not investigated in this work. The velocity is replaced using the relations $\bm{n}\cdot\bm{v}=\dot{r}$, $v^2=\dot{r}^2 + r^2\dot{\phi}^2$ and $\bm{n}\times\bm{v}=r\dot{\phi}\bm{\ell}$.  Then, by a PN iteration, these expressions are inverted to obtain $\dot{r}^2$ and $\dot{\phi}$ is terms of the energy~$E$, the angular momentum~$J$, and the orbital radius $r$. The results exactly match the structure of \eqref{eq:rdot2phidot}, so the coefficients entering \eqref{eq:RS} are immediately identified.  In order to present these results nicely, it is useful to introduce the dimensionless reduced energy and angular momentum~\cite{Blanchet:2013haa}, which read
\be\label{eq:def_epsilon_j} \varepsilon = -\frac{2E}{\mu c^2} \qquad\quad\mathrm{and}\quad\qquad j = - \frac{2 J^2 E}{\mu^3(\tilde{G}\alpha m)^2}\,.\ee

Note that $\varepsilon = \calO(c^{-2})$ but $j = \calO(1)$.  The desired coefficients (given in machine-readable form in the Supplementary Material~\cite{SuppMaterial}) then read 
\begin{widetext}
\bse\label{eq:expression_ABCDFI_ST}\begin{align}
A &= - c^2 \varepsilon \left\{1 +  \frac{3}{4} \varepsilon \left[1 - 3 \nu\right] + \frac{1}{8} \varepsilon^2 \left[4 - 19 \nu + 16 \nu^2\right]+ \mathcal{O}\left(\varepsilon^3\right) \right\}\\
B&=  \tilde{G} \alpha m  \left\{1 + \varepsilon \left[3 + \bar{\gamma} -  \frac{7}{2} \nu\right] + \varepsilon^2 \left[\frac{9}{4} + \frac{3}{4} \bar{\gamma} + \nu\Big(-12 -  \frac{15}{4} \bar{\gamma}\Big) + \frac{21}{4} \nu^2\right] + \mathcal{O}\left(\varepsilon^3\right)  \right\}\\
C &=- \frac{(\tilde{G}\alpha m)^2 j}{c^2 \varepsilon} \Bigg\{1 + \varepsilon \left[1  - 3 \nu + \frac{10 + 2 \bar{\beta}_{+} + 4 \bar{\gamma} - 2 \bar{\beta}_{-} \delta - 5 \nu}{j}\right] \nn\\
& \quad  + \varepsilon^2 \left[\frac{3-15\nu+15\nu^4}{4}  + \frac{18 + 2 \bar{\beta}_{+} + 12 \bar{\gamma} + 2 \bar{\gamma}^2 - 2 \bar{\beta}_{-} \delta + \nu\Big(- \frac{127}{2} - 4 \bar{\beta}_{+} - 26 \bar{\gamma} + 6 \bar{\beta}_{-} \delta\Big)  + 18 \nu^2}{j}\right] + \mathcal{O}\left(\varepsilon^3\right)  \Bigg\}\\
D_1&= \frac{(\tilde{G}\alpha m)^3 j}{c^4 \varepsilon} \Bigg\{8 + 4 \bar{\gamma} - 3 \nu \nn\\
& \qquad\qquad+ \varepsilon \Bigg[8 + 4 \bar{\gamma} + \nu\Big(- \frac{61}{2} - 13 \bar{\gamma}\Big) + \frac{19}{2} \nu^2 \nn\\
&  \qquad\qquad\qquad+ \frac{1}{j}\Bigg(26 + 12 \bar{\beta}_{+} + \frac{2}{3} \bar{\delta}_{+} + \frac{62}{3} \bar{\gamma} + 4 \bar{\beta}_{+} \bar{\gamma} + \frac{25}{6} \bar{\gamma}^2 -  \frac{4}{3} \bar{\chi}_{+} + \delta\Big(-12 \bar{\beta}_{-} + \frac{2}{3} \bar{\delta}_{-} - 4 \bar{\beta}_{-} \bar{\gamma} + \frac{4}{3} \bar{\chi}_{-}\Big)  \nn\\
&\qquad\qquad\qquad\qquad+ \nu\Big(- \frac{81}{2} -  \frac{4}{3} \bar{\delta}_{+} -  \frac{58}{3} \bar{\gamma} -  \frac{1}{3} \bar{\gamma}^2 + 16 \bar{\beta}_{-}^2 \bar{\gamma}^{-1} - 16 \bar{\beta}_{+}^2 \bar{\gamma}^{-1} + \frac{8}{3} \bar{\chi}_{+} + 8 \bar{\beta}_{-} \delta\Big)  + 10 \nu^2\Bigg)\Bigg]+ \mathcal{O}\left(\varepsilon^2\right) \Bigg\}\\
D_2&=\frac{(\tilde{G}\alpha m)^4 j}{c^6 \varepsilon}\left\{-33 -  \bar{\delta}_{+} - 33 \bar{\gamma} -  \frac{33}{4} \bar{\gamma}^2 -  \bar{\delta}_{-} \delta + \nu \Big(\frac{75}{4} + 2 \bar{\beta}_{+} + 8 \bar{\gamma} - 2 \bar{\beta}_{-} \delta\Big)  - 6 \nu^2 + \mathcal{O}\left(\varepsilon\right)  \right\} \\
D_3&= \frac{(\tilde{G}\alpha m)^5 j^2 \nu}{c^8 \varepsilon^2} \left\{\frac{15}{4} + 2 \bar{\gamma} -  \frac{1}{4} \nu  + \mathcal{O}\left(\varepsilon\right)  \right\}\\
F &= \frac{\tilde{G}\alpha m }{c}\sqrt{\frac{j}{\varepsilon}} \left\{1 + \frac{1-3\nu}{2}\varepsilon +\frac{2-9\nu+6\nu^2}{8} \varepsilon^2  + \mathcal{O}\left(\varepsilon^3\right)  \right\}\\
I_1&= - \frac{(\tilde{G}\alpha m)^2 }{c^3} \sqrt{\frac{j}{\varepsilon}} \bigg\{4 + 2 \bar{\gamma} - 2 \nu + \varepsilon \big[2 +  \bar{\gamma}+ \nu(-10 - 4 \bar{\gamma})  + 4 \nu^2\big]  + \mathcal{O}\left(\varepsilon^2\right)  \bigg\}\\
I_2& = \frac{(\tilde{G}\alpha m)^3}{c^5}\sqrt{\frac{j}{\varepsilon}} \left\{9 + \bar{\delta}_{+} + 9 \bar{\gamma} + \frac{9}{4} \bar{\gamma}^2 + \bar{\delta}_{-} \delta +\nu  \Big(- \frac{3}{4} - 4 \bar{\beta}_{+} + 2 \bar{\gamma} + 2 \bar{\beta}_{-} \delta\Big)  + 5 \nu^2  + \mathcal{O}\left(\varepsilon\right)  \right\}\\
I_3&= - \frac{ (\tilde{G}\alpha m)^4\nu}{c^7}\left(\frac{j}{\varepsilon}\right)^{3/2}\left\{\frac{3}{2} + \bar{\gamma}  +  \nu  + \mathcal{O}\left(\varepsilon\right)  \right\}
\end{align}\ese
\end{widetext}
Plugging these coefficients into \eqref{eq:n_K_general} and \eqref{eq:PK_parameters_general}, it is straightforward to obtain the expressions of the 2PK parameters in terms of $\varepsilon$ and $j$, and of course the parameters of the ST theory. However, the final expression are quite lengthy, so I have relegated them to Appendix~\ref{app:PKparameters}.

\section{The flux of energy and angular momentum}
\label{sec:flux}
As an application of the QK parametrization obtained above, I will now derive in Section~\ref{subsec:flux_res} the energy and angular momentum flux for eccentric orbits at Newtonian order, i.e. at next-to-leading order beyond the leading dipolar radiation. Before this, it is necessary to derive in Section~\ref{subsec:flux_gen_results} some generic formulas for the angular momentum flux associated to the scalar field.

\subsection{General expressions for the energy and angular momentum fluxes}
\label{subsec:flux_gen_results}
In GR, the conservation of energy and angular momentum lead to balance equations of the Bondi energy $E$ and angular momentum $S_i$ of the form \mbox{$\dd E/\dd t = - \mathcal{F}$} and \mbox{$\dd S_i/\dd t = - \mathcal{G}_ i$}, where $\mathcal{F}$ and $\mathcal{G}_ i$ are, respectively, the energy and angular momentum fluxes. These are very useful for the determination of the evolution of the orbital parameters under radiation reaction. In the PN-MPM formalism, the fluxes entering these balance equations are deduced thanks to the Gauss theorem and from the divergenceless of the Landau-Lifschitz stress-energy pseudotensor, namely $\p_\mu \tau^{\mu\nu} = 0$, which reduced to $\p_\mu \Lambda^{\mu\nu}=0$ in the vacuum zone exterior to the matter. It is thus sufficient to compute the fluxes of energy and angular momentum through an infinitely large sphere centered around the position of the CM, and generic formulas for this are given by the right-hand sides of (4.4) of~\cite{Blanchet:2018yqa}. In the Einstein frame, such formulas are directly applicable to the $\tau^{\mu\nu}$ of ST theories as given by \eqref{eq:taumunu}, and the fluxes neatly subdivide into a tensor and scalar sector, namely $\dd E/\dd t = - \mathcal{F}- \mathcal{F}^s$ and $\dd S_i/\dd t = - \mathcal{G}_ i - \mathcal{G}^s_ i$. The tensor contributions $\mathcal{F}$ and $\mathcal{G}^i$ are formally given by the same expressions as in GR~\cite{Thorne:1980ru,Blanchet:2013haa} up to a factor $\phi_0$, but beware that the definition of $h^{\mu\nu}$ in GR and ST theories actually differ. The tensor sector being fully treated, I will only focus on the scalar sector in this section. Expressions for the scalar energy flux are already known~\cite{Dgef92,Lang:2014osa,Will:2018bme}, but analogous expressions for the angular momentum have not yet been calculated~\cite{Will:2018bme}, although very similar expressions have been derived in a different setup~\cite{Zhang:2018prg,Khalil:2022sii}.

Starting from (4.4) of~\cite{Blanchet:2018yqa}, I find that the desired quantities read
\bse\label{eq:scalar_fluxes_1}\begin{align}
 \mathcal{F}^s &= \lim_{R\rightarrow \infty}\frac{c^5 \phi_0 R^2}{16\pi G} \int \dd^2 \Omega \,N_k \Lambda_\phi^{0k}(R,\mathbf{N}) \,, \\
\mathcal{G}_i^s &= \lim_{R\rightarrow \infty}\frac{c^4 \phi_0 R^3}{16\pi G} \epsilon_{iab} \int \dd^2 \Omega \, N_k N_b \Lambda_\phi^{ka}(R,\mathbf{N})  \,,
\end{align}\ese
where $\int \dd^2 \Omega$ means integration over angles. Thanks to the $R\rightarrow 0$ limit, only the asymptotic expression of $\Lambda_\phi$ is needed, namely
\be \label{eq:Lambda_phi_asymptotic} \Lambda_\phi^{\mu\nu} = (3+2\omega_0)\big(\eta^{\mu\alpha}\eta^{\nu\beta}- \frac{1}{2} \eta^{\mu\nu}\eta^{\alpha\beta}\big)\p_\alpha \psi \p_\beta \psi + \calO\left(R^{-3}\right) . \ee

Plugging this expression into \eqref{eq:scalar_fluxes_1}, I obtain\footnote{Following Section 11.5.2 of~\cite{Will:2018bme}, I actually needed to use the structure of $\psi = \sum_\ell f_\ell(T-R/c)\hat{N}_L/R + \calO(R^{-2})$, where $f_\ell$ are some arbitrary functions. One can then state \mbox{$\p_k \psi = (-N_k/c)\p_t \psi + \calO(R^{-2})$}.
Moreover, the convergence of the $R\rightarrow \infty$ limit for the angular momentum is ensured by observing that \mbox{$\epsilon_{iab} N_b \p_a \psi = \epsilon_{iab}N_b(-N_a/c)\p_t\psi + \calO(R^{-2}) = \calO(R^{-2})$}.}:
\bse \label{eq:scalar_fluxes_2}\begin{align}
\mathcal{F}^s &= \frac{c^3 R^2(3+2\omega_0)\phi_0}{16\pi G} \int \dd^2 \Omega \,\dot{\psi}^2 \label{seq:scalar_energy_flux_2} \\
\mathcal{G}^s_i &= \frac{c^3 R^3 (3+2\omega_0) \phi_0}{16 \pi G}\int \dd^2 \Omega\,  \dot{\psi}\,\epsilon_{iab} N_a \p_b\psi  \label{seq:scalar_angular_momentum_flux_2}
\end{align}\ese
 The expression for \eqref{seq:scalar_energy_flux_2} was already known [see (6.6) of~\cite{Lang:2014osa}], but \eqref{seq:scalar_angular_momentum_flux_2} is new, and agrees up to a global prefactor with (3.4) of~\cite{Zhang:2018prg} and (B13) of~\cite{Khalil:2022sii}, which were derived in a different setup.

Replacing $\psi$ by its multipolar expansion given in~\eqref{seq:scalar_radiative_expansion} leads to the final expression:\footnote{The $\mathcal{O}(1/R^2)$ subleading terms of the asymptotic multipolar expansion are discarded, see \cite{Compere:2019gft} for a justification within the Bondi-Sachs framework.}
\bse \label{eq:scalar_fluxes_3}\begin{align}
\mathcal{F}^s &= \sum_{\ell=0}^{\infty} \frac{G \phi_0 (3+2\omega_0)}{c^{2\ell+1} \ell! (2\ell+1)!!}\overset{(1)}{\mathcal{U}}{}^s_L\overset{(1)}{\mathcal{U}}{}^s_L \label{seq:scalar_energy_flux_3} \,,\\&\nn\\
\mathcal{G}^s_i &= \sum_{\ell=1}^{\infty} \frac{G \phi_0 (3+2\omega_0)}{c^{2\ell+1} (\ell-1)! (2\ell+1)!!}\,\epsilon_{iab}\,\mathcal{U}^s_{aL-1} \overset{(1)}{\mathcal{U}}{}^s_{bL-1}  	  \label{seq:scalar_angular_momentum_flux_3}\,.
\end{align}\ese

\subsection{Expressions for the fluxes at next-to-leading (Newtonian) order}
\label{subsec:flux_res}
This subsection is devoted to the computation of the flux of energy and angular momentum at Newtonian order beyond standard quadrupolar radiation, which means next-to-leading order beyond the leading order dipolar radiation of ST theories. Although in theory, one has access to the 1.5PN fluxes using the PK parametrization developed in Section~\ref{sec:PK}, I chose to limit myself to Newtonian order because hereditary integrals enter as soon as 0.5PN in ST theories, see \citetalias{Bernard:2022noq} (or~\cite{Lang:2014osa}). The latter integrals can in general only be treated analytically in the $e\rightarrow 0$ limit, and even in this case, they are tedious to compute, so I keep these for future work. This restriction also allows me to present results that are valid for any eccentricity~$e$. 

At Newtonian order, the nonlinearities of the MPM construction do not play a role and the fluxes can be simply expressed in terms of the source moments introduced in Section~\ref{sec:STtheory}. The fluxes read (neglecting $\mathcal{O}(c^{-3})$ terms):
\bse \label{eq:fluxes_N}\begin{align}
\mathcal{F} &= 	\frac{G\phi_0}{5c^{5}}\overset{(3)}{\mathrm{I}}_{\!\!ij}\overset{(3)}{\mathrm{I}}_{\!\!ij}  \label{seq:tensor_energy_flux_N} \,,\\
\mathcal{G}_i &=  \frac{2G\phi_0}{5c^{5}}\epsilon_{iab}\overset{(2)}{\mathrm{I}}_{\!\!ak}\overset{(3)}{\mathrm{I}}_{\!\!bk}  \label{seq:tensor_angular_momentum_flux_N}\,,\\
\mathcal{F}^s &=  \frac{G \phi_0 (3+2\omega_0)}{c^3}\Bigg[ \frac{\overset{(2)}{\mathrm{I}}{}^{\!\!s}_{\!\!a}\overset{(2)}{\mathrm{I}}{}^{\!\!s}_{\!\!a} }{3}+ \frac{\overset{(1)}{E}{}^{\!s}\overset{(1)}{E}{}^{\!s}}{\phi_0^2 c^2} + \frac{\overset{(3)}{\mathrm{I}}{}^{\!\!s}_{\!\!ab}\overset{(3)}{\mathrm{I}}{}^{\!\!s}_{\!\!ab} }{30c^2} \Bigg]  \label{seq:scalar_energy_flux_N} \,,\\
\mathcal{G}^s_i &= \frac{G \phi_0 (3+2\omega_0)}{c^3}\epsilon_{iab}\Bigg[\frac{1}{3}\overset{(1)}{\mathrm{I}}{}^{\!\!s}_{\!\!a}\overset{(2)}{\mathrm{I}}{}^{\!\!s}_{\!\!b} +  \frac{1}{15c^2}\overset{(2)}{\mathrm{I}}{}^{\!\!s}_{\!\!ak}\overset{(3)}{\mathrm{I}}{}^{\!\!s}_{\!\!bk}  \Bigg] \,.
\end{align}\ese

The required moments are given explicitly in (4.6) and (4.7) of \citetalias{Bernard:2022noq}, while the acceleration needed to compute the time derivatives is given by   (3.10) of~\cite{Bernard:2018ivi} (but see also~\cite{Mirshekari:2013vb}). Since  only  nonspinning particles are considered, there is no precession of the orbital plane, so it is possible to write $\mathcal{G}_i = \mathcal{G} \ell_i$ and $\mathcal{G}^s_i = \mathcal{G}^s \ell_i$, where $\bm{\ell}$ is the constant unit vector orthogonal to the orbital plane. 

For eccentric orbits, the fluxes will undergo small modulations with frequencies comparable to an orbital period due to the presence of trigonometric functions of the eccentric anomaly $u$, and are presented explicitly in machine-readable form in the Supplementary Material~\cite{SuppMaterial}. In order to obtain the secular evolution of the orbital parameters, it is very useful to instead consider the orbit-averaged fluxes. Since one is averaging over a \emph{radial} period $P$ between two passages through the periastron, the orbit averaging procedure reads~\cite{Blanchet:2013haa} $$\left\langle\mathcal{\mathcal{H}} \right\rangle \equiv \frac{1}{P}\int_0^P \dd t \, \mathcal{H}(t) = \frac{1}{2\pi}\int_0^{2\pi}\dd u\, \frac{\dd\ell}{\dd u} \mathcal{H}(u)\,,$$ where $\mathcal{H}$ stands for either $\mathcal{F}$, $\mathcal{F}^s$, $\mathcal{G}$ or $\mathcal{G}^s$, and $\dd\ell/\dd u$ is given at this order by 
\be \frac{\dd \ell}{\dd u} = 1-e_t \cos u  + \calO(x^2) \,.\ee
The time-averaging procedure naturally involves computing integrals of the type
$$\frac{1}{2\pi} \int_{0}^{2\pi}\dd u\,  \frac{\cos^p(u) \sin^q(u)}{\left[1-e_t \cos(u)\right]^n} \,, $$
but with basic algebra and trigonometry, these can be reduced to the following two families of integrals whose values are well known for $n\ge 1$~\cite{BS89}:
\begin{align}
\frac{1}{2\pi} \int_{0}^{2\pi} \!\!\frac{\dd u}{\left[1-e_t \cos(u)\right]^n} &= \frac{1}{(1-e_t^2)^{n/2}}P_{n-1}\!\left(\!\frac{1}{\sqrt{1-e_t^2}}\!\right)\,,\\*
\frac{1}{2\pi} \int_{0}^{2\pi} \!\!\frac{\dd u \,\sin u}{\left[1-e_t \cos(u)\right]^n} &= 0\,,
\end{align}
where $P_n(x)$ is the usual Legendre polynomial of order~$n$ (the $n\le 0$ case is not required and is trivial to compute anyway).

Finally, all results will be expressed only in terms of the quasi-invariant quantity \mbox{$x = (\tilde{G}\alpha mKn/c^3)^{2/3}$} and the gauge-dependent time eccentricity $e_t$. The reason for this is that it allows to immediately recover the usual expressions for circular orbits by taking $e_t \rightarrow 0$. Since all the PK parameters are expressed in \eqref{eq:expression_ABCDFI_ST} in terms of $\varepsilon$ and $j$, it is necessary to invert \eqref{eq:PK_parameters_expressions_ST} at 1PN order to obtain
\begin{align}\label{eq:varepsilon_j_in_terms_of_x_et}
\varepsilon &= x + \frac{x^2}{12(1-e_t^2)}\bigg(-9+8 \bar\beta_+ - 8 \bar\gamma - 8 \bar\beta_- \delta - \nu \nn\\*
&\qquad\qquad\qquad\qquad\quad+ e_t^2 \Big[-15-8\bar\gamma+\nu\Big]\bigg) + \mathcal{O}(x^3)\,, \\
j &= 1 - e_t^2 + \frac{x}{4} \bigg(9-8 \bar\beta_+ + 8 \bar\gamma + 8 \bar\beta_- \delta + \nu \nn\\*
&\qquad\qquad\qquad\qquad\quad+ e_t^2 \Big[-17-8\bar\gamma+7\nu\Big]\bigg) + \mathcal{O}(x^2) \,.
\end{align}

Note that it is also possible to express everything in terms of the two quasi-invariants quantities $(x,\iota)$ where \mbox{$\iota \equiv (3 + 2 \overline{\gamma} -\overline{\beta}_{+} +   \overline{\beta}_{-} \delta)x/k$} reduces to $j$ in the $x\rightarrow 0$ limit~\cite{Blanchet:2013haa}. The latter choice however, prohibits us from reading off the circular-orbit limit directly from the results, which is why I did not use it here. The time-averaged fluxes (given in machine-readable form in the Supplementary Material~\cite{SuppMaterial}) then read at Newtonian order:
 \begin{subequations}\label{eq:orbitAverageFlux_tensor}
\begin{align}
\langle\mathcal{F} \rangle &= \frac{32 c^5 x^5 \nu^2 (1+\bar{\gamma}/2)}{5\tilde{G} \alpha}\cdot \frac{1 + \frac{73}{24}e_t^2  + \frac{37}{96} e_t^4}{(1-e_t^2)^{7/2}}\,,\\
\langle\mathcal{G} \rangle &= \frac{32 c^2 m x^{7/2} \nu^2 (1+\bar{\gamma}/2)}{5}\cdot\frac{1 + \frac{7}{8}e_t^2 }{(1-e_t^2)^{2}}\,,
\end{align}
\end{subequations}
 \begin{widetext}
 \begin{subequations}\label{eq:orbitAverageFlux_scalar}
\begin{align}
\langle\mathcal{F}^s \rangle &= \frac{c^5 x^5  \nu^2  \zeta }{3\tilde{G} \alpha} \Bigg(4  \mathcal{S}_{-}^2 x^{-1} \cdot \frac{1 +  \frac{1}{2}e_t^2}{(1-e_t^2)^{5/2}} \nn \\
& \qquad\qquad +\frac{1}{15(1-e_t^2)^{7/2}} \Bigg\{-24 \zeta^{-1} \bar{\gamma} - 120 \mathcal{S}_{-}^2 - 80 \bar{\beta}_{+} \mathcal{S}_{-}^2  + 240 \bar{\beta}_{+} \bar{\gamma}^{-1} \mathcal{S}_{-}^2 - 40 \bar{\gamma} \mathcal{S}_{-}^2 + 240 \bar{\beta}_{-} \bar{\gamma}^{-1} \mathcal{S}_{-} \mathcal{S}_{+}\nn\\
&\qquad\qquad\qquad\qquad\qquad\qquad  + \delta \Big[80 \bar{\beta}_{-} \mathcal{S}_{-}^2 - 240 \bar{\beta}_{-} \bar{\gamma}^{-1} \mathcal{S}_{-}^2 - 240 \bar{\beta}_{+} \bar{\gamma}^{-1} \mathcal{S}_{-} \mathcal{S}_{+}\Big]  - 80 \mathcal{S}_{-}^2 \nu \nn\\
&\qquad\qquad\qquad\qquad\quad  + e_t^2 \Bigg[480 \zeta^{-1} \bar\beta_+ - 720 \zeta^{-1}\bar\gamma^{-1} \bar\beta_+^2 - 153 \zeta^{-1} \bar\gamma + 1080 \mathcal{S}_-^2 - 280  \bar\beta_+ \mathcal{S}_-^2  + 1440  \bar\gamma^{-2} \bar\beta_-^2  \mathcal{S}_-^2 \nn\\
 &\qquad\qquad\qquad\qquad\qquad\quad + 1440  \bar\gamma^{-2} \bar\beta_+^2  \mathcal{S}_-^2  - 240 \bar\gamma^{-1} \bar\beta_+ \mathcal{S}_-^2 + 540  \bar\gamma \mathcal{S}_-^2 + 2880 \bar\gamma^{-2} \bar\beta_- \bar\beta_+ \mathcal{S}_-\mathcal{S}_+ - 240 \bar\gamma^{-1} \bar\beta_-\mathcal{S}_- \mathcal{S}_+  \nn\\
&\qquad\qquad\qquad\qquad\qquad\quad  + \delta \Big[280 \bar\beta_- \mathcal{S}_-^2 - 480 \bar\gamma^{-1}\bar\beta_- \mathcal{S}_-^2 - 480 \bar\gamma^{-1} \bar\beta_+ \mathcal{S}_-\mathcal{S}_+\Big] - 620 \mathcal{S}_-^2 \nu   \Bigg] 
\Bigg\}  \nn\\
&\qquad\qquad\qquad\qquad\quad  + e_t^4 \Bigg[120 \zeta^{-1} \bar\beta_+ - 180 \zeta^{-1}\bar\gamma^{-1} \bar\beta_+^2  - \frac{117}{4}\zeta^{-1}\bar\gamma + 195 \mathcal{S}_-^2 + 360 \bar\gamma^{-2}\bar\beta_-^2  \mathcal{S}_-^2 + 360 \bar\gamma^{-2}\beta_+^2  \mathcal{S}_-^2  \nn\\
 &\qquad\qquad\qquad\qquad\qquad\quad  - 150 \bar\gamma^{-1} \bar\beta_+ \mathcal{S}_-^2  + 70 \bar\gamma \mathcal{S}_-^2 + 720 \bar\gamma^{-2} \bar\beta_-  \bar\beta_+\mathcal{S}_- \mathcal{S}_+ - 150 \bar\gamma^{-1} \bar\beta_- \mathcal{S}_- \mathcal{S}_+ \nn\\
&\qquad\qquad\qquad\qquad\qquad\quad  + \delta \Big[-30 \bar\gamma^{-1} \bar\beta_- \mathcal{S}_-^2 - 30 \bar\gamma^{-1} \bar\beta_+ \mathcal{S}_- \mathcal{S}_+\Big] -125 \mathcal{S}_-^2  \nu   \Bigg] 
\Bigg\}  
\Bigg)\,,\\
\langle\mathcal{G}^s \rangle &= \frac{c^2 m x^{7/2}  \nu^2  \zeta }{3} \Bigg(\frac{4  \mathcal{S}_{-}^2 x^{-1}}{1-e_t^2} \nn \\
& \qquad\qquad +\frac{1}{15(1-e_t^2)^{2}} \Bigg\{-24 \zeta^{-1} \bar{\gamma} - 120 \mathcal{S}_{-}^2 - 80 \bar{\beta}_{+} \mathcal{S}_{-}^2 + 240 \bar{\beta}_{+} \bar{\gamma}^{-1} \mathcal{S}_{-}^2 - 40 \bar{\gamma} \mathcal{S}_{-}^2  + 240 \bar{\beta}_{-} \bar{\gamma}^{-1} \mathcal{S}_{-} \mathcal{S}_{+}\nn\\
&\qquad\qquad\qquad\qquad\qquad\qquad  + \delta \Big[80 \bar{\beta}_{-} \mathcal{S}_{-}^2 - 240 \bar{\beta}_{-} \bar{\gamma}^{-1} \mathcal{S}_{-}^2 - 240 \bar{\beta}_{+} \bar{\gamma}^{-1} \mathcal{S}_{-} \mathcal{S}_{+}\Big]  - 80 \mathcal{S}_{-}^2 \nu \nn\\
&\qquad\qquad\qquad\qquad\quad  + e_t^2 \Bigg[- 21 \zeta^{-1}\bar\gamma  + 390 \mathcal{S}_-^2 - 60 \bar\beta_+\mathcal{S}_-^2 + 120 \bar\gamma^{-1}\bar\beta_+\mathcal{S}_-^2 + 200 \bar\gamma \mathcal{S}_-^2 + 120\bar\gamma^{-1}\bar\beta_-\mathcal{S}_-\mathcal{S}_+   \nn\\
&\qquad\qquad\qquad\qquad\qquad\quad  + \delta \Big[60 \bar\beta_- \mathcal{S}_-^2 - 120\bar\gamma^{-1} \bar\beta_- \mathcal{S}_-^2 - 120 \bar\gamma^{-1} \bar\beta_+\mathcal{S}_-\mathcal{S}_+\Big] -  220 \mathcal{S}_-^2\nu   \Bigg] 
\Bigg\}  \Bigg)\,.
\end{align}
\end{subequations}
\end{widetext}

It is easy to verify that in the limit of circular orbits $e_t \rightarrow 0$, the energy fluxes coincide with the results for circular orbits of \citetalias{Bernard:2022noq}. Moreover, I verified that for circular orbits, the angular momentum fluxes indeed verify the well-known relations $\mathcal{F}_\mathrm{circ} = \omega \,\mathcal{G}_\mathrm{circ}$ and $\mathcal{F}^s_\mathrm{circ} = \omega \,\mathcal{G}^s_\mathrm{circ}$, where $\omega = Kn ={c^3 x^{3/2}}/(\tilde{G}\alpha m)$. Finally, $\left\langle \mathcal{F} \right\rangle$ and $\left\langle\mathcal{G}\right\rangle$ yield the correct expressions in the GR limit~\cite{PM63}, while $\left\langle\mathcal{F}^s\right\rangle$ and $\left\langle\mathcal{G}^s\right\rangle$ vanish.

\section{Evolution of the orbital elements}

Thanks to both the QK parametrization of Section~\ref{sec:PK} and the fluxes obtained in Section~\ref{sec:flux}, it is now possible to obtain from flux-balance arguments the secular evolution of the PK parameters under the effect of radiation reaction. The method is briefly sketched in Section~\ref{subsec:orbital_elements_gen}, and applied at lowest $-1$PN order in Section~\ref{subsec:Peters_Mathews}. At this order, it is possible to analytically solve the ODEs arising from the flux-balance equations and thus obtain expressions analogous to those of Peters and Mathews~\cite{PM63,Peters64} for GR. The relative 1PN expressions for the ODEs governing the orbital evolution are then presented in Section~\ref{subsec:orbital_elements_1PN}.

\label{sec:orbital_elements}
\subsection{General method}
\label{subsec:orbital_elements_gen}
Consider an orbital element, which is denoted generically $\xi$. Its orbit-averaged evolution can be expressed as 
\be \label{eq:secular_evolution_orbital_parameters} \left\langle \frac{\dd \xi}{\dd t}\right\rangle = \frac{\p \xi}{\p E} \left\langle \frac{\dd E}{\dd t}\right\rangle  + \frac{\p \xi}{\p J} \left\langle  \frac{\dd J}{\dd t}\right\rangle  \,. \ee
where $J=|\mathbf{J}|$.
Since the expressions of the orbital elements in terms of $E$ and $J$ were computed in Section \ref{sec:PK} and the energy and angular momentum fluxes were computed in Section \ref{sec:flux}, we now have all the ingredients to compute the expressions of the orbital elements, assuming that the following balance equations hold:
\be
 \label{eq:orbit_average_balance_equations}
   \left\langle \frac{\dd E}{\dd t}\right\rangle  = -\left\langle \mathcal{F}\right\rangle-\left\langle \mathcal{F}^s\right\rangle    \quad \mathrm{and}\quad \left\langle  \frac{\dd J}{\dd t}\right\rangle = - \left\langle \mathcal{G}\right\rangle-\left\langle \mathcal{G}^s\right\rangle \,.
\ee

\subsection{``Peters and Mathews'' formulas for ST theory}

\label{subsec:Peters_Mathews}

First, let us work at lowest Newtonian order in the orbital parameters, and reproduce in this case some results derived in GR by Peters and Mathews~\cite{PM63,Peters64}. At this order, there is no difference between the difference eccentricities, which is simply denoted in this subsection as $e$ (instead of $e_r$, $e_t$ and $e_\phi$). Similarly, the semimajor axis will be denoted in this subsection~\mbox{$a$ (instead of $a_r$),} which is related at this order to the variable $x$ by \mbox{$x = \tilde{G}\alpha m/(c^2 a)$}. The secular evolution of $a$ and $e$ is driven only by the leading order, $-1$PN scalar fluxes of energy and angular momentum, and one easily derives from \eqref{eq:secular_evolution_orbital_parameters} the following equations:
\bse\begin{align}
\left\langle\frac{\dd a}{\dd t}\right\rangle &= -  \frac{8}{3} \,\frac{(\tilde{G}\alpha m)^2 \zeta \mathcal{S}_-^2 \nu}{c^3}\cdot\frac{1+e^2/2}{a^2(1-e^2)^{5/2}}\label{seq:dedt_N_1}\, \\
\left\langle\frac{\dd e}{\dd t}\right\rangle  &= - \frac{(\tilde{G}\alpha m)^2 \zeta\mathcal{S}_-^2 \nu}{c^3}\cdot \frac{2e}{a^3 (1-e^2)^{3/2}} \label{seq:dedt_N_1}\,.
\end{align}\ese
These expressions have the same dependencies in~$a$ and~$e$ as (4.5) and (4.6) of~\cite{Zhang:2018prg}, but the coefficients are tedious to relate due to different initial formulations. It is thus immediately to  find (now dropping the time-averaging brackets)
\be \frac{\dd a}{\dd e} = \frac{4a(1+e^2/2)}{3e(1-e^2)}\ee
This is straightforward to integrate, and it finally leads to
\be \label{eq:PetersMathews_ST}a(e) = \frac{c_0 e^{4/3}}{1-e^2}\,,\ee
where $c_0$ is an integration constant determined only by the initial condition $(a_0,e_0)$. This is the ST equivalent of the famous Eq. (5.11) of~\cite{Peters64} in GR, which I reproduce here for clarity:
\be \label{eq:PetersMathews_GR}a_\mathrm{GR}(e) = \frac{c_0^\mathrm{GR} e^{12/19}}{1-e^2}\left(1+\frac{121}{304}e^2\right)^{870/2299}\,.\ee
I would like to insist that \eqref{eq:PetersMathews_ST} assumes that the system is only driven by the leading-order dipolar terms in ST theories, neglecting quadrupolar subleading terms, while \eqref{eq:PetersMathews_GR} arises purely from the leading-order quadrupolar radiation in GR. Quite remarkably, \eqref{eq:PetersMathews_ST} does not depend on any free parameter of ST theory, but this statement heavily relies on the assumption that the radiation is dipole driven. Unlike its GR counterpart, it is analytically invertible, and its inverse is given by
\bse\be
 e(a) = \sqrt{1 - \frac{c_0^3}{3 a^3}- \frac{c_0^3(6a^3-c_0^3)}{3 a^3\sqrt[3]{\Xi}} +  \frac{\sqrt[3]{\Xi}}{3 a^3} }\,,
\ee
 where
 \be \Xi = - \frac{27}{2} a^6 c_0^3 + 9 a^3 c_0^6 - c_0^9 + \frac{3}{2}\sqrt{3 a^9 c_0^6 (27a^3-4c_0^3)}\,.\ee
 \ese

\begin{figure}
\includegraphics[width=0.4\textwidth]{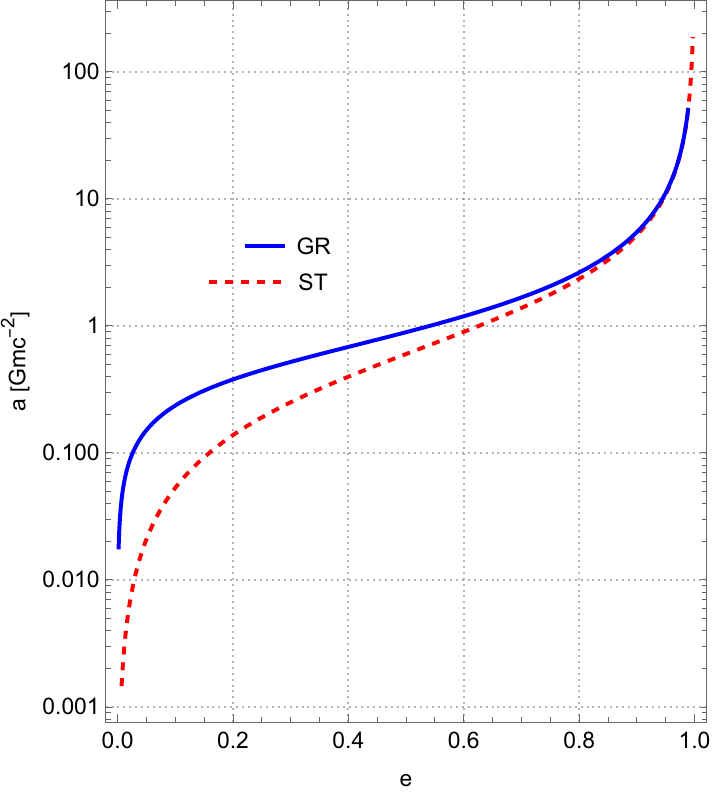}\caption{\label{fig:a(e)}Plot of $a(e)$ at leading order in GR and ST, with geometrical units ($G=c=m=1$). The constant $c_0$ is chosen to be $c_0=1$ in GR and $c_0=(425/304)^{870/2299}\approx 1.135$ in ST theory, such that the asymptotic expressions of $a(e)$ when $e\rightarrow 1$ coincide in both theories.}
\end{figure}

\begin{figure}
\includegraphics[width=0.4\textwidth,trim={0 1.26cm 0 0},clip]{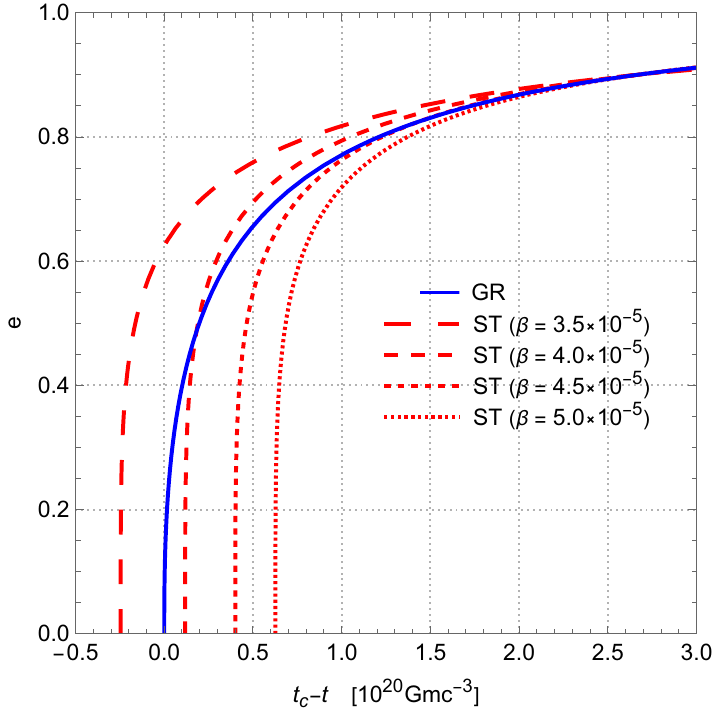}
\includegraphics[width=0.4\textwidth]{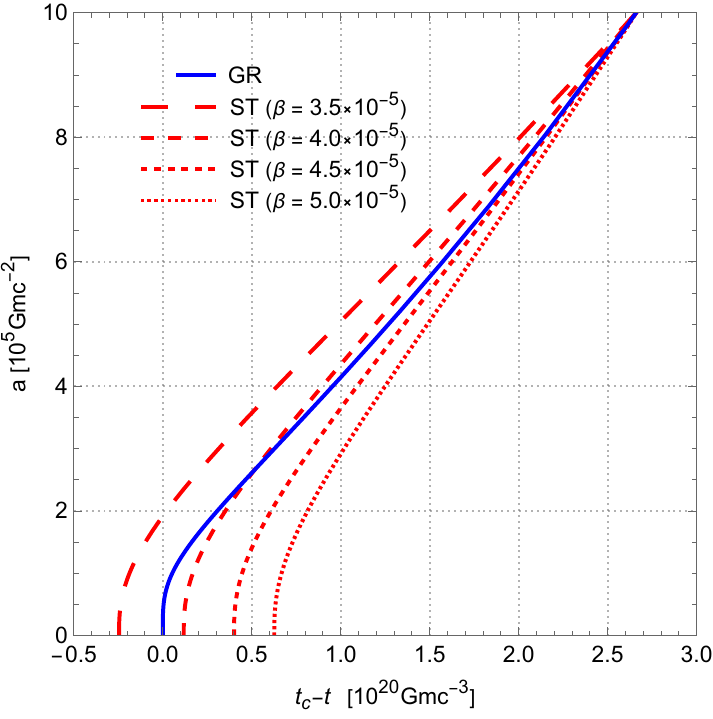}
\caption{\label{fig:e(t)a(t)} Time evolution for an equal-mass binary at leading order of $e(t)$ and $a(t)$ in GR and ST tensor theories for several values of the parameter $\beta$ (use geometrical units $G=c=m=1$). Time is expressed as the time remaining until coalescence \emph{in the GR case}. The initial condition are set such that in all theories, the binary's  has an eccentricity $e=0.9$ and a semimajor axis \mbox{$a_{0.9}=10^6$} at some given time $t_{0.9}$ (here, \mbox{$t_c-t_{0.9} \approx 2.659$}). In the ST case $\beta = 4.0\times10^{-5}$, the binary circularizes slower than in GR at earlier times \mbox{$t_c-t_{0.9}>t_c-t>2\times10^{19}$}, and faster than GR at later times \mbox{$t_c-t<2\times10^{19}$}.}
\end{figure}

 In order to compare the two functions, switch to geometrical units $G=c=m=1$ for the rest of the subsection. I follow~\cite{Peters64} in choosing $c_0^\mathrm{GR} = 1$ in the GR case, and then choose the ST value for $c_0$ such that $a(e)$ and $a_\mathrm{GR}(e)$ have the same asymptotic behavior as $e \rightarrow1$. This turns out to be \mbox{$c_0 = (425/304)^\frac{870}{2299}\approx 1.135$}. The two functions $a(e)$ and $a_\mathrm{GR}(e)$ are compared in Fig.~\ref{fig:a(e)}.

One can now plug in the expression \eqref{eq:PetersMathews_ST} for $a(e)$ into~\eqref{seq:dedt_N_1}, and which leads to an expression analogous to (5.13) of~\cite{Peters64} which here reads
  \be \left\langle\frac{\dd e}{\dd t}\right\rangle  = - \frac{3\beta(1-e^2)^{3/2}}{4c_0^3 e^3} \,,\ee
 where $$\beta=  \frac{8}{3}\frac{(\tilde{G}\alpha m)^2 \zeta\mathcal{S}_-^2 \nu}{c^3} \,.$$
Unlike in GR
, this ODE is straightforward to integrate, leading to
\be \label{eq:t_in_terms_of_e_N}t_c - t = \frac{8 c_0^3}{3\beta}\left[\frac{1-e^2/2}{\sqrt{1-e^2}} - 1 \right] > 0\,,\ee
where $t_c$ is an arbitrary constant which is chosen to be the instant of coalescence, such that \mbox{$a(t) \rightarrow 0$} and \mbox{$e(t) \rightarrow 0$} when \mbox{$t \rightarrow t_c^+$}.

This expression is yet again invertable, and its inverse reads
\be e(t) = \sqrt{2(\varrho+1)\sqrt{\varrho(\varrho+2)}-4\varrho-2\varrho^2}\,,\ee
where $\varrho \equiv 3\beta(t_c-t)/(8 c_0^3)>0$. Plugging this expression into \eqref{eq:PetersMathews_ST} directly yields $a(t)$.

 To compare the time evolution of eccentricity in GR and ST, let us plot \eqref{eq:t_in_terms_of_e_N} against the numerically integrated solution of (5.13) of~\cite{Peters64}. Consider the case of an equal mass binary, i.e. $\nu=1/4$. Thus, the ``$\beta$'' defined in~\cite{Peters64} is given $\beta_\mathrm{GR} = 16/5$, while in ST theory $\beta$ is the only free parameter of the theory at this order. Finally, I set up the initial condition by imposing that for each theory, there is a common time $t_{0.9}$ where the binary has eccentricity $e=0.9$ and semimajor axis $a_{0.9}=10^6$. The time evolution is plotted in Fig.~\ref{fig:e(t)a(t)}. The behavior depends strongly on the choice for the parameter $\beta$, but for certain regimes, the binary initially circularizes slower than in GR, but after some given time,  it instead circularizes faster.

\subsection{Time evolution of $x$ and $e_t$}
\label{subsec:orbital_elements_1PN}
At next-to-leading, Newtonian order, one cannot obtain closed form expressions for the time evolution of the orbital elements. In order to avoid clutter, I only present hereafter the evolution of the two main variables used to describe the motion, $x$ and $e_t$ (they are also given in machine-readable form in the Supplementary Material~\cite{SuppMaterial}). Thanks to the relations displayed in Appendix \ref{app:PKparameters} along with \eqref{eq:varepsilon_j_in_terms_of_x_et}, the reader can then obtain the secular time evolution of all the QK orbital elements by simple algebraic replacements. The evolution equations read

\bse\label{eq:ODEs_x_et}\begin{align}
\label{seq:ODE_x}\left\langle\frac{\dd x}{\dd t} \right\rangle&=\frac{2 c^3 \zeta x^4 \nu}{3  \tilde{G} \alpha m} \Bigg\{\frac{4  \mathcal{S}_{-}^2(1 + \frac{1}{2} e_t^2)}{(1 -  e_t^2)^{5/2}} \nn\\
&+ \frac{x}{15 (1 -  e_t^2)^{7/2}} \Big(\mathcal{C}_1 + e_t^2 \mathcal{C}_2 + e_t^4 \mathcal{C}_3  \Big) + \mathcal{O}(x^2)\Bigg\}\,, \\
\label{seq:ODE_et}\left\langle\frac{\dd e_t}{\dd t}\right\rangle &=- \frac{c^3 \zeta x^3 \nu}{ \tilde{G} \alpha m} \Bigg\{\frac{2 \mathcal{S}_{-}^2e_t }{(1 -  e_t^2)^{3/2}} \nn\\*
&\qquad\quad+ \frac{x \,e_t}{15 (1 -  e_t^2)^{5/2}} \Big(\mathcal{C}_4 +e_t^2 \mathcal{C}_5  \Big) + \mathcal{O}(x^2) \Bigg\}\,,
\end{align}\ese
where the constants $\mathcal{C}_n$ depend only on the masses and the scalar-tensor parameters, and read

\begin{widetext}
\bse\begin{align}
\mathcal{C}_1 &= 288 \bar{\zeta}^{-1} + 120 \bar{\zeta}^{-1} \bar{\gamma} - 30 \mathcal{S}_{-}^2 - 160 \bar{\beta}_{+} \mathcal{S}_{-}^2 + 240 \bar{\beta}_{+} \bar{\gamma}^{-1} \mathcal{S}_{-}^2 + 40 \bar{\gamma} \mathcal{S}_{-}^2  + 240 \bar{\beta}_{-} \bar{\gamma}^{-1} \mathcal{S}_{-} \mathcal{S}_{+} \nn\\*
&\quad +  \delta\Big(160 \bar{\beta}_{-} \mathcal{S}_{-}^2 - 240 \bar{\beta}_{-} \bar{\gamma}^{-1} \mathcal{S}_{-}^2 - 240 \bar{\beta}_{+} \bar{\gamma}^{-1} \mathcal{S}_{-} \mathcal{S}_{+}\Big) - 70 \mathcal{S}_{-}^2 \nu \,, \\
\mathcal{C}_2 &= 876 \bar{\zeta}^{-1} + 480 \bar{\beta}_{+} \bar{\zeta}^{-1} + 285 \bar{\zeta}^{-1} \bar{\gamma} - 720 \bar{\beta}_{+}^2 \bar{\zeta}^{-1} \bar{\gamma}^{-1} + 1095 \mathcal{S}_{-}^2 - 260 \bar{\beta}_{+} \mathcal{S}_{-}^2 + 540 \bar{\gamma} \mathcal{S}_{-}^2 - 240 \bar{\beta}_{+} \bar{\gamma}^{-1} \mathcal{S}_{-}^2 \nn\\*
&\quad + 1440 \bar{\beta}_{-}^2 \bar{\gamma}^{-2} \mathcal{S}_{-}^2 + 1440 \bar{\beta}_{+}^2 \bar{\gamma}^{-2} \mathcal{S}_{-}^2 - 240 \bar{\beta}_{-} \bar{\gamma}^{-1} \mathcal{S}_{-} \mathcal{S}_{+} + 2880 \bar{\beta}_{-} \bar{\beta}_{+} \bar{\gamma}^{-2} \mathcal{S}_{-} \mathcal{S}_{+}  \nn\\*
&\quad +\delta \Big(260 \bar{\beta}_{-} \mathcal{S}_{-}^2 - 480 \bar{\beta}_{-} \bar{\gamma}^{-1} \mathcal{S}_{-}^2 - 480 \bar{\beta}_{+} \bar{\gamma}^{-1} \mathcal{S}_{-} \mathcal{S}_{+}\Big)  - 625 \mathcal{S}_{-}^2 \nu \, ,\\
\mathcal{C}_3 &= 111 \bar{\zeta}^{-1} + 120 \bar{\beta}_{+} \bar{\zeta}^{-1} + \frac{105}{4} \bar{\zeta}^{-1} \bar{\gamma} - 180 \bar{\beta}_{+}^2 \bar{\zeta}^{-1} \bar{\gamma}^{-1} + 270 \mathcal{S}_{-}^2 + 110 \bar{\gamma} \mathcal{S}_{-}^2 - 150 \bar{\beta}_{+} \bar{\gamma}^{-1} \mathcal{S}_{-}^2 + 360 \bar{\beta}_{-}^2 \bar{\gamma}^{-2} \mathcal{S}_{-}^2 \nn\\*
&\quad+ 360 \bar{\beta}_{+}^2 \bar{\gamma}^{-2} \mathcal{S}_{-}^2 - 150 \bar{\beta}_{-} \bar{\gamma}^{-1} \mathcal{S}_{-} \mathcal{S}_{+} + 720 \bar{\beta}_{-} \bar{\beta}_{+} \bar{\gamma}^{-2} \mathcal{S}_{-} \mathcal{S}_{+}+ \delta\Big(-30 \bar{\beta}_{-} \bar{\gamma}^{-1} \mathcal{S}_{-}^2 - 30 \bar{\beta}_{+} \bar{\gamma}^{-1} \mathcal{S}_{-} \mathcal{S}_{+}\Big) - 130 \mathcal{S}_{-}^2  \nu \,, \\
\mathcal{C}_4 &= 304 \bar{\zeta}^{-1} + 160 \bar{\beta}_{+} \bar{\zeta}^{-1} + 100 \bar{\zeta}^{-1} \bar{\gamma} - 240 \bar{\beta}_{+}^2 \bar{\zeta}^{-1} \bar{\gamma}^{-1} + 165 \mathcal{S}_{-}^2 - 120 \bar{\beta}_{+} \mathcal{S}_{-}^2 + 100 \bar{\gamma} \mathcal{S}_{-}^2 - 40 \bar{\beta}_{+} \bar{\gamma}^{-1} \mathcal{S}_{-}^2 + 480 \bar{\beta}_{-}^2 \bar{\gamma}^{-2} \mathcal{S}_{-}^2\nn\\*
&\quad + 480 \bar{\beta}_{+}^2 \bar{\gamma}^{-2} \mathcal{S}_{-}^2 - 40 \bar{\beta}_{-} \bar{\gamma}^{-1} \mathcal{S}_{-} \mathcal{S}_{+} + 960 \bar{\beta}_{-} \bar{\beta}_{+} \bar{\gamma}^{-2} \mathcal{S}_{-} \mathcal{S}_{+} \nn\\*
&\quad + \delta\Big(120 \bar{\beta}_{-} \mathcal{S}_{-}^2 - 200 \bar{\beta}_{-} \bar{\gamma}^{-1} \mathcal{S}_{-}^2 - 200 \bar{\beta}_{+} \bar{\gamma}^{-1} \mathcal{S}_{-} \mathcal{S}_{+}\Big)  - 125 \mathcal{S}_{-}^2 \nu \,,  \\
\mathcal{C}_5 &= 121 \bar{\zeta}^{-1} + 40 \bar{\beta}_{+} \bar{\zeta}^{-1} + \frac{175}{4} \bar{\zeta}^{-1} \bar{\gamma} - 60 \bar{\beta}_{+}^2 \bar{\zeta}^{-1} \bar{\gamma}^{-1} + 280 \mathcal{S}_{-}^2 - 20 \bar{\beta}_{+} \mathcal{S}_{-}^2 + 130 \bar{\gamma} \mathcal{S}_{-}^2 - 10 \bar{\beta}_{+} \bar{\gamma}^{-1} \mathcal{S}_{-}^2 + 120 \bar{\beta}_{-}^2 \bar{\gamma}^{-2} \mathcal{S}_{-}^2 \nn\\*
&\quad + 120 \bar{\beta}_{+}^2 \bar{\gamma}^{-2} \mathcal{S}_{-}^2 - 10 \bar{\beta}_{-} \bar{\gamma}^{-1} \mathcal{S}_{-} \mathcal{S}_{+} + 240 \bar{\beta}_{-} \bar{\beta}_{+} \bar{\gamma}^{-2} \mathcal{S}_{-} \mathcal{S}_{+} \nn\\*
&\quad+ \delta\Big(20 \bar{\beta}_{-} \mathcal{S}_{-}^2 - 50 \bar{\beta}_{-} \bar{\gamma}^{-1} \mathcal{S}_{-}^2 - 50 \bar{\beta}_{+} \bar{\gamma}^{-1} \mathcal{S}_{-} \mathcal{S}_{+}\Big) - 150 \mathcal{S}_{-}^2 \nu \, .
\end{align}\ese


Solving these ODEs numerically 
yields the secularly evolving orbital elements $x(t)$ and $e_t(t)$, which can be replaced into all results derived in terms of $x$ and $e_t$.
\newpage\end{widetext}

\section{Waveform}
\label{sec:waveform}
Another important application of the QK parametrization is the computation of gravitational wave templates. To this end, I will now present the waveform  for eccentric orbits decomposed into spherical harmonic modes. As in \citetalias{Bernard:2022noq}, the gothic \emph{conformal} metric is decomposed into two independent modes along the polarization vectors,
\begin{subequations}
\begin{align}
h_+& \equiv \frac{1}{2}(P_i P_j - Q_i Q_j) h_{ij}^\mathrm{TT}\,,\\
h_\times & \equiv \frac{1}{2}(P_i Q_j + Q_i P_j) h_{ij}^\mathrm{TT}\,,
\end{align}
\end{subequations}
which can be recast into a complex field, $h = h_+ - \di h_\times$. The latter can be decomposed on the basis on spin-weighted spherical harmonics of weight $-2$,
\begin{subequations}
\be h = h_+ - \di h_\times = \sum_{\ell=2}^{+\infty} \sum_{m=-\ell}^\ell h^{\ell m} \,_ {-2}Y^{\ell m}(\Theta,\Phi)\,.\ee
Similarly, the pure spin-0 scalar field can be decomposed on standard (spin-0) spherical harmonics,
\be \psi =\sum_{\ell=0}^{+\infty} \sum_{m=-\ell}^\ell \psi^{\ell m}\  Y^{\ell m}(\Theta,\Phi) \,.\ee
\end{subequations}
The modes are in practice computed using the relations
\begin{subequations}
\begin{align}
\label{hlmUlmVlm}
h^{\ell m} &= - \frac{G}{\sqrt{2} R c^{\ell+2}}\Big[ \mathcal{U}^{\ell m} - \frac{\di}{c} \mathcal{V}^{\ell m}\Big]\,,\\
\psi^{\ell m} &= \frac{G}{R c^{\ell+2}} \mathcal{U}_s^{\ell m}\,,
\end{align}
\end{subequations}
where
\begin{subequations}
\begin{align}
\mathcal{U}^{\ell m} &= \frac{4}{\ell!}\sqrt{\frac{(\ell+1)(\ell+2)}{2\ell(\ell-1)}}\alpha_L^{\ell m} \mathcal{U}_L\,,\\
\mathcal{V}^{\ell m} &= -\frac{8}{\ell!}\sqrt{\frac{\ell(\ell+2)}{2(\ell+1)(\ell-1)}}\alpha_L^{\ell m} \mathcal{V}_L\,,\\
\mathcal{U}_s^{\ell m} &= -\frac{2}{\ell!} \alpha_L^{\ell m} \mathcal{U}^s_L\,,
\end{align}
\end{subequations}
and where $\alpha_L^{\ell m}$ is defined by\footnote{See \citetalias{Bernard:2022noq} for an explicit expression.}
\begin{align}
\alpha_L^{\ell m} &\equiv \int \dd \Omega \,\hat{N}_L {\left(Y^{\ell m}\right)}^*
\end{align}

 These modes can then be recast into dimensionless amplitude modes $h^{\ell m}$ and $\psi^{\ell m}$ which are given by\footnote{
 There is no need to introduce an auxiliary phase as in \citetalias{Bernard:2022noq} at this order.}
\begin{subequations}
\begin{align}
h^{\ell m} &= \frac{2 \tilde{G}(1-\zeta) m \nu x}{R c^2}  \sqrt{\frac{16\pi}{5}} \,\hat{H}^{\ell m} \de^{-\di \dm \phi } \,,\\
\psi^{\ell m} &= \frac{2 i \tilde{G} \zeta \sqrt{\alpha}\mathcal{S}_{-} m \nu \sqrt{x}}{R c^2} \sqrt{\frac{8\pi}{3}} \,\hat{\Psi}^{\ell m} \de^{-\di \dm \phi}\,.
\end{align}
\end{subequations}
with normalized modes defined such that the expression for circular orbits reads \mbox{$\hat{H}^{22}_\mathrm{circ} = 1 + \mathcal{O}(x)$} and \mbox{$\hat{\Psi}^{11}_\mathrm{circ} = 1 + \mathcal{O}({x})$}. Recalling that $h^{\ell, -m} = (-)^\ell {\left(h^{\ell m}\right)}^*$ and $\psi^{\ell, -m} = (-)^\ell {\left(\psi^{\ell m}\right)}^*$,  only the $m \ge 0$ modes that enter the Newtonian (next-to-leading) waveform are presented. In the eccentric case, they exhibit  time oscillations which enter via the eccentric anomaly~$u$, see the Kepler equation \eqref{seq:kepler_equation}. They are given in machine-readable form in the Supplementary Material~\cite{SuppMaterial}, and read

\bse\label{eq:modes}\begin{align}
\hat{\Psi}^{00} &= -\frac{\di}{\sqrt{x}\mathcal{S}_-}\sqrt{\frac{3}{2}}\Bigg\{\frac{\mathcal{S}_+ +  \mathcal{S}_- \delta}{\nu} + \frac{x}{1-e_t \cos u}\nn\\*
&\qquad\quad\times\bigg[8 \bar\beta_- \bar\gamma^{-1} \mathcal{S}_-  - \frac{5}{2}\mathcal{S}_+ + 8 \bar\beta_+ \bar\gamma^{-1} \mathcal{S}_+ + \frac{1}{2}\mathcal{S}_- \delta \nn\\*
&\qquad\qquad\qquad-\frac{e_t \cos u}{6}\Big(\mathcal{S}_+ -  \mathcal{S}_- \delta\Big)\bigg] \Bigg\}+ \mathcal{O}(x) \\
\hat{\Psi}^{11} &= \frac{1-e_t^2 - \di e_t \sqrt{1-e_t^2}\sin u}{\sqrt{1-e_t^2}(1-e_t\cos u)}+ \mathcal{O}(x)\\
\hat{\Psi}^{22} &= \frac{\di \sqrt{x}\left(\mathcal{S}_+ - \mathcal{S}_- \delta\right)}{\sqrt{5}\mathcal{S}_- (1-\cos u)^2}\Bigg\{1- \frac{1}{2}e_t \cos u  \nn\\*
&\quad+ \frac{1}{4}e_t^2\left(-3+\cos 2u\right) - \di e_t\sqrt{1-e_t^2}\sin u\Bigg\}+ \mathcal{O}(x)\\
\hat{\Psi}^{20} &= \frac{\di \sqrt{x} e_t \cos u \left(\mathcal{S}_+ - \mathcal{S}_- \delta\right)}{\sqrt{30}\mathcal{S}_- (1-e_t\cos u)}\\
\hat{H}^{22} &= \frac{1}{(1-e_t \cos u)^2}\Bigg\{1- \frac{1}{2}e_t \cos u \nn\\*
&\quad+ \frac{1}{4} e_t^2 \left(-3+\cos 2u\right) - \di e_t\sqrt{1-e_t^2}\sin u\Bigg\}+ \mathcal{O}(x)\\
\hat{H}^{20} &= \frac{e_t \cos u}{\sqrt{6}(1-e_t\cos u)}+ \mathcal{O}(x)
\end{align}\ese

These results agree in the GR limit with~\cite{Mishra:2015bqa}, as well as with \citetalias{Bernard:2022noq} in the circular $e_t \rightarrow 0$ limit.

\section{Conclusion}

To my knowledge, this work is the first systematic analytical study of eccentric binaries in ST theories. The expressions of the 2PK orbital elements were computed, first in a generic setting, then specializing to ST theories. The flux of energy and angular momentum at next-to-leading order beyond dipolar radiation were then derived, i.e. at Newtonian order with respect to quadrupolar radiation. After a preliminary study of the leading order $-1$PN case, for which were obtained several closed form expressions analogous to the works of Peters and Mathews~\cite{PM63,Peters64},  the energy and angular momentum balance equations were then used to obtain the ODEs expressing the secular evolution of the orbital elements at Newtonian order. Finally, the expressions for waveform in terms of spherical harmonic modes were computed at Newtonian order. At all steps of these computations, I checked that these results were in agreement with the literature in the GR limit as well in limiting case of circular orbits ($e_t \rightarrow 0$).

Thanks to the 2PK parametrization computed in this work, it is in principle possible to obtain the fluxes, waveform and evolution of the orbital elements at 1.5PN order, just like in \citetalias{Bernard:2022noq}. However, one difficulty for this is the treatment of the hereditary tail and memory integrals in the eccentric case. In GR, it is possible to treat these analytically~\cite{ABIQ08tail,ABIQ08,ABIS09} as an expansion in eccentricity for $e_t \ll 1$, and the generalization to the case of ST would not present any additional difficulty. Another possible extension of this work is the derivation of the 3PK parametrization. The main difficulty would be the treatment of the nonlocal tails in the equations and conserved energy. Extending the techniques developed for GR at 4PN order to treat these hereditary terms~\cite{ChoTanay2022} seems possible as well.

Finally, I would like to emphasize that the results of Section~\ref{subsec:PK_general} are not theory dependent: they can be used as soon as one obtains an expression for the energy and angular momentum that exhibits the right properties. These results are to be used as a facilitator in studying the behavior of compact binary systems on eccentric orbits for a range of different theories of gravity. 
\acknowledgments
The author thanks Laura Bernard, Luc Blanchet and Sashwat Tanay and  for useful discussions and comments on the final manuscript, as well as the Institut d'Astrophysique de Paris for its hospitality. Most computations were performed using the xAct library~\cite{xtensor} for \emph{Mathematica}~\cite{Mathematica}. The author received support from the Czech Academy of Sciences under the Grant No. LQ100102101.
Finally, the author thanks Davide Usseglio, Abhishek Chowdhuri and Tom Colin for noticing some typos in the published version. 

\appendix
\section{Integration formulas using complex analysis}\label{app:complex}

The goal of this section is to sketch the computation of integrals of the type \eqref{eq:P_Phi_integrals}. It was proven that the PN expansion of these integrals can readily be performed by integrating the PN expansion of the integrand~\cite{DS88}. Since $D_1$, $D_2$ and $D_3$ are all small in the PN sense, with the help of \eqref{eq:R_factorized}, both integrals of \eqref{eq:P_Phi_integrals} can be written in the form
\be\sum_{k\in\mathbb{N}} \int_{s_a}^{s_p} \frac{\dd s \, \overline{P}_k(s)}{s^2 \Big\{A+ 2 B  s+C s^2\Big\}^{k/2}} \,,\label{eq:structure_PN_expanded_integral}\ee
where $\overline{P}_k(s)$ is a polynomial that arises from the PN expansion of the integrand. This naturally leads to posing the master integral\footnote{The case $q=0$ is considered because it appears in the computation of the radial action, see (3.5) of~\cite{DS88}.}
\be \mathcal{I}_{p,q} = \frac{1}{\pi} \int_{s_a}^{s_p} \dd s\, s^{p-2} \left(A+ 2B s + C s^2\right)^{1/2-q} \,,\ee
where $(p,q) \in \mathbb{N}^2$, and recall that $A<0$, $B>0$, and \mbox{$C<0$}. Following a method initially found by Sommerfeld~\cite{Sommerfeld}, this integral is astutely promoted into a path integral in the complex plane~\cite{Golstein1980, DS88}:
\be \label{eq:master_integral_complex}\mathcal{I}_{p,q} = \frac{1}{2\pi} \oint_\mathcal{C} \dd s\, s^{p-2} \left(A + 2B s + C s^2 \right)^{1/2-q} \,.\ee

\begin{figure}
\includegraphics[width=0.45\textwidth]{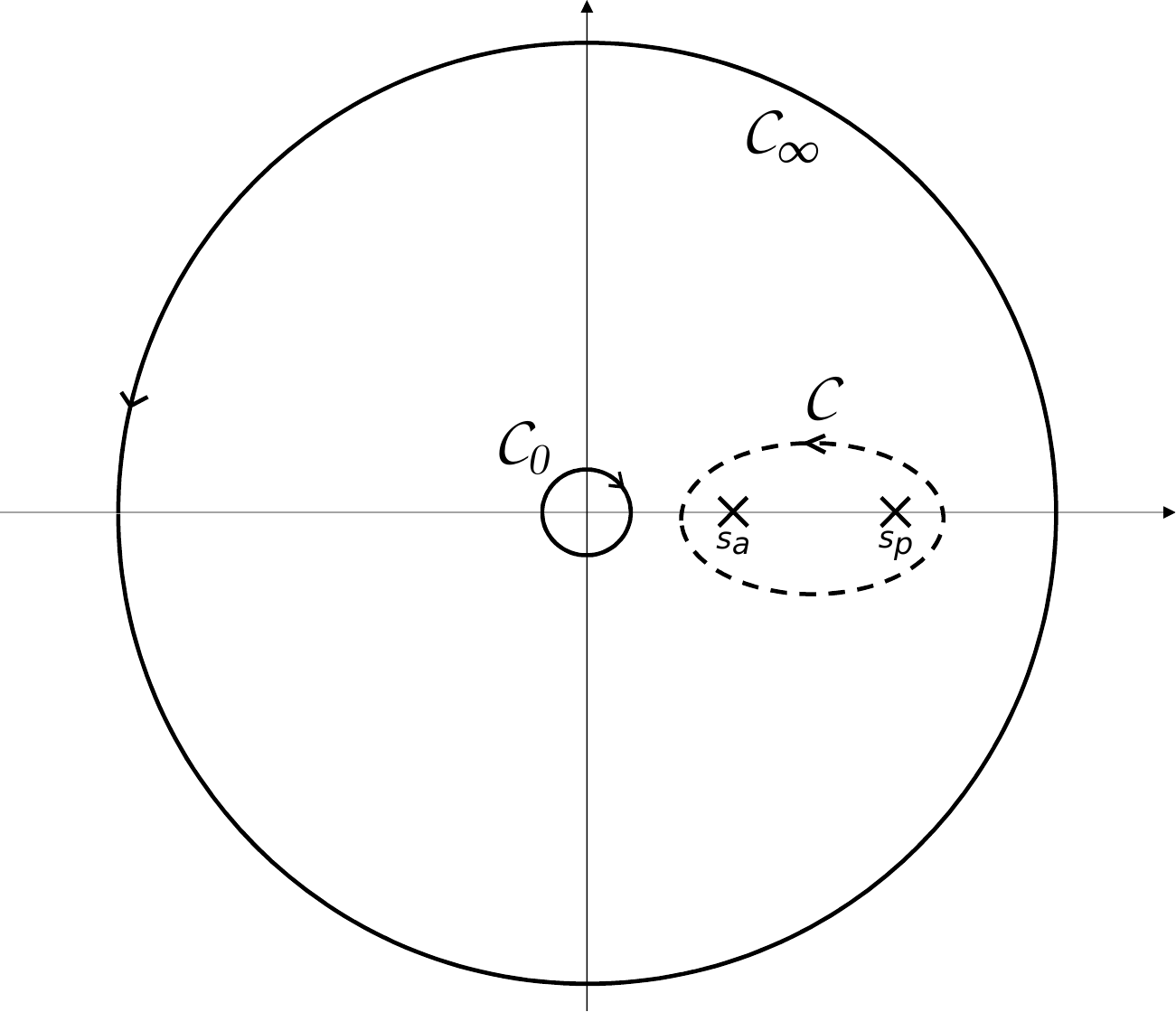}
	\caption{\label{fig:contour}Sketch of the integration contours $\mathcal{C}$, $\mathcal{C}_0$ and $\mathcal{C}_\infty$, adapted from~\cite{BBBFMb}. The integrand of \eqref{eq:master_integral_complex} exhibits poles at $0$ and $\infty$, and is meromorphic everywhere except in the neighborhood of the segment~$[s_a, s_p]$. }
\end{figure}
The contour $\mathcal{C}$ loops  counterclockwise around the two roots $s_a$ and $s_p$, avoiding the poles at $s=0$, see Fig.~\ref{fig:contour}. The integrand is not meromorphic inside of $\mathcal{C}$, so the residue theorem cannot be directly applied. However, it is possible to deform the contour into $\mathcal{C}_0$ and $\mathcal{C}_\infty$, which, respectively contain the pole at~$0$ and~$\infty$ (the pole at infinity is mapped back to $0$ thanks to the change of variables $r = 1/s$). The integrands are now meromorphic within the contours, so the residue theorem is applied. The end result reads

\begin{align}
\mathcal{I}_{p,q} &= \big[p=0\big](-1)^q(1-2q)B(-A)^{-1/2-q}\nn \\*
& + \big[p=1\big] (-1)^{q+1} (-A)^{1/2-q} \nn\\*
& + \big[p\ge 2q\big](-1)^{q+1}(-C)^{1/2-q}\nn\\*
&\!\!\!\!\!\times\!\!\!\!\!\!\sum_{k=\ceil{\frac{p}{2}-q}}^{p-2q}\!\! \frac{\Gamma\left(\frac{1}{2}-q+1\right) A^{p-k-2q}\, (2B)^{2k+2q-p}\,C^{-k}}{\Gamma\left(\frac{1}{2}-q-k+1\right) (2k+2q-p)!(p-k-2q)!}\,,
 \end{align}
 where $\Gamma(x)$ is the usual gamma function, $\ceil{n}$ is the ceiling of the integer $n$, and $\big[\mathcal{P}\big]$ is the Inverson bracket defined such that $[\mathcal{P}] = 1$ if $\mathcal{P}$ is true, and $[\mathcal{P}] = 0$ otherwise. When applied to (3.8) of~\cite{DS88}, this formula yields indeed the correct result, given by (3.9) of~\cite{DS88}.

\section{Details of the quasi-Keplerian construction at 2PN order}\label{app:PKconstruction}

The starting point of the determination of the QK parametrization is \eqref{eq:rdot2phidot}. The periods $P$ and $K$ defined by \eqref{eq:P_Phi_integrals} are obtained using the techniques of Appendix~\ref{app:complex}, from which one trivially obtains  the mean motion $n = 2\pi/P$. The positions of the periastron $r_p$ and apastron $r_a$ are first obtained at lowest order by factorizing the $c\rightarrow \infty$ limit of $\mathcal{R}(s)$. PN corrections to these roots are  then obtained iteratively by requiring that they cancel $\mathcal{R}(s)$, up to higher-order PN corrections. The PK parameters \mbox{$a_r = (s_p+s_a)/(2s_a s_p)$} and \mbox{$e_r = (s_p-s_a)/(s_p+s_a)$} are then trivial to obtain, where we recall $s_{p,a}=1/r_{p,a}$.
Following~\cite{MGS04}, it is now very useful to introduce the auxiliary variable

\be \tilde{v} \equiv 2 \arctan \left(\sqrt{\frac{1+e_r}{1-e_r}} \tan\left(\frac{u}{2}\right)\right) \,, \label{eq:vt}\ee
which differs from $v$ defined in \eqref{eq:v} by the choice of eccentricity. Note that at this stage, $e_\phi$ and $e_t$ are undetermined, only $e_r$ is known.

The starting point to determine the Kepler equation is to integrate \eqref{seq:rdot2} and find
\be \label{seq:time_integral}   \left| t-t_0 \right|  = \int_{s}^{s_p} \frac{\dd s}{s^2\sqrt{\mathcal{R}(s)}} = \int_{s}^{s_p} \frac{\dd s\, \overline{P}(s)}{s^2 \sqrt{(s-s_a)(s_p-s)} } \,,  \ee
where $t_0$ is some reference time at which the system is an periastron and where $\overline{P}(s)$ arises [similarly to \eqref{eq:structure_PN_expanded_integral}] from the PN expansion of $1/\sqrt{\widetilde{\mathcal{R}}\smash{(s)}}$, see \eqref{eq:R_factorized}. The problem has now been reduced to computing master integrals of the form

\be \mathcal{J}_p \equiv  \int^{s} \frac{\dd s\, s^{p-2}}{\sqrt{(s-s_a)(s_p-s)}} \,, \ee

Interestingly, these master integrals are most conveniently expressed using auxiliary variables that depend heavily on the value of $p$. They are given by 
\bse \label{eq:J_master_integrals}
\begin{align}
\mathcal{J}_0 &= \frac{(s_a+ s_p) u - (s_p-s_a)\sin(u)}{2 (s_a s_p)^{3/2}}\\
\mathcal{J}_1 &= \frac{u}{\sqrt{s_a s_p}}\\
\mathcal{J}_2 &= \tilde{v} \\
\mathcal{J}_3 &= \frac{(s_a+s_p)\tilde{v}+(s_p - s_a)  \sin(\tilde{v})}{2} \\
\mathcal{J}_4 &= \frac{3s_a^2+2s_a s_p + 3 s_p^2}{8} \tilde{v} + \frac{(s_p-s_a)(s_p+s_a)}{2}\sin(\tilde{v}) \nn\\*
&\quad+ \frac{(s_p-s_a)^2}{16}\sin(2\tilde{v}) \\
\mathcal{J}_5 &= \frac{(s_p+s_a)(5s_a^2 -2s_a s_p+5s_p^2)}{16}\tilde{v} \nn\\*
&\quad+ \frac{3(s_p - s_a)(5s_a^2+6s_as_p+5s_p^2)}{32}\sin(\tilde{v}) \nn\\*
&\quad + \frac{3(s_p-s_a)^2(s_a+s_p)}{32}\sin(2\tilde{v})\nn \\*
&\quad + \frac{(s_p-s_a)^3}{96}\sin(3\tilde{v})
\end{align}
\ese

Keeping with the notation of~\cite{MGS04}, and replacing $s_p$ and~$s_a$ by their 2PN accurate expressions, I explicitly find  an intermediate form for the Kepler equation,
\be  n(t-t_0) = u + \kappa_0 \sin(u) + \frac{\kappa_1}{c^4}(\tilde{v}-u) + \frac{\kappa_2}{c^4}\sin(\tilde{v}) \,.  \label{seq:kepler_equation_vt}\ee

Similarly, the computation of the angular equation starts by integrating \eqref{seq:phidot} to find
\be \label{seq:time_integral}   \left| \phi-\phi_0 \right| =  \int_{s}^{s_p} \frac{\dd s \,\mathcal{S}(s)}{s^2\sqrt{\mathcal{R}(s)}} =  \int_{s}^{s_p} \frac{\dd s\, \overline{Q}(s)}{s^2 \sqrt{(s-s_a)(s_p-s)} } \,,  \ee
where $\phi_0$ is the reference angle at $t_0$ and where $\overline{Q}(s)$ arises from the PN expansion of $\mathcal{S}(s)/\sqrt{\tilde{R}\smash{(s)}}$. The master integrals of \eqref{eq:J_master_integrals} apply, and we find the explicit intermediate form of the angular equation to be

\be  \frac{2\pi}{\Phi}\left(\phi - \phi_0 \right) = \tilde{v} + \frac{\lambda_1}{c^2} \sin(\tilde{v}) + \frac{\lambda_2}{c^4}\sin(2\tilde{v}) + \frac{\lambda_3}{c^4}\sin(3\tilde{v})  \,. \label{seq:angular_equation_vt}\ee

Let us now turn to the problem of expressing the Kepler and angular equations in terms of $v$. First, parametrize the 2PN-accurate relation between~$e_r$ and~$e_\phi$~as
\be e_r = e_\phi\left(1+ \frac{\alpha}{c^2} + \frac{\beta}{c^4}\right)  \,, \label{eq:er_to_ephi_alpha_beta}\ee
and Taylor-expand Eq. \eqref{eq:vt}. To this end, first write
\be \sqrt{\frac{1+e_r}{1-e_r}} = \sqrt{\frac{1+e_\phi}{1-e_\phi}}\left(1+ \frac{\kappa}{c^2}\right) \,,\ee
where 
\be \kappa = \frac{\alpha e_\phi}{(1-e_\phi^2)} + \frac{e_\phi (2\beta + \alpha^2 e_\phi + 2 e_\phi^2 (\alpha^2 - \beta)}{2c^2(1-e_\phi^2)^2} +\calO\left(\frac{1}{c^4}\right)\,.\ee

Then, thanks to the identities
\bse\begin{align}
\arctan'(x) &=\frac{\sin(2\arctan(x))}{2x}\,,\\
\arctan''(x) &=  \frac{\sin(4 \arctan(x))-2\sin(2\arctan(x))}{4 x^2}\,,
\end{align}\ese
the relation between $\tilde{v}$ and $v$ can then be written
\be \tilde{v}=v+\frac{\kappa}{c^2} \sin(v) + \frac{\kappa^2}{4c^4}\left(\sin(2v)-2\sin(v)\right)+\calO\left(\frac{1}{c^6}\right)\,. \label{eq:relation_vt_v}\ee
Moreover, the latter relation immediately leads to
\be \sin(\tilde{v}) = \sin(v) + \frac{\kappa}{2c^2} \sin(2v) + \calO\left(\frac{1}{c^4}\right) \label{eq:relation_sinvt_sinv} \,. \ee
Replacing these relations into the intermediate angular relation \eqref{seq:angular_equation_vt}, one obtains an equation with the same form as  \eqref{seq:angular_equation}, \emph{except} that there is an extra term proportional to $\sin(v)$. Now is the time to use the freedom in $\alpha$ and $\beta$ to set the coefficient in front of $\sin(v)$  to zero. In this way, the relation \eqref{eq:er_to_ephi_alpha_beta} between $e_\phi$ and $e_r$ is entirely determined. Finally, since the intermediate Kepler equation~\eqref{seq:kepler_equation_vt} only features $\tilde{v}$ at 2PN order, one can simply replace it by $v$ so as to obtain exactly the form of \eqref{seq:kepler_equation}. It is then straightforward to simply read off all the PK parameters.

\begin{widetext}
\section{Expressions for post-Keplerian parameters in massless scalar-tensor theories}\label{app:PKparameters}
The post-Keplerian parameters in terms of $\varepsilon$, $j$ and the ST parameters are given in machine-readable form in the Supplementary Material~\cite{SuppMaterial}, and are presented hereafter as well.
\bse\label{eq:PK_parameters_expressions_ST}\begin{align}
n &=\frac{c^3\varepsilon^{3/2} }{ \tilde{G}\alpha m} \Bigg\{1 + \varepsilon \Bigg[- \frac{15}{8} -  \bar{\gamma} + \frac{\nu}{8} \Bigg] \nn\\*
&\qquad+ \varepsilon^2 \Bigg[\frac{555}{128} + \frac{33}{8} \bar{\gamma} + \bar{\gamma}^2 +\nu \left(\frac{15}{64} + \frac{1}{8}\bar{\gamma} \right)  + \frac{11}{128} \nu^2 +  \frac{-15(1+\bar\gamma+\frac{\bar\gamma^2}{4}) +  \bar{\delta}_{+}  +   \bar{\delta}_{-} \delta +  \nu\Big(6 - 2 \bar{\beta}_{+} + 4 \bar{\gamma} + 2\bar{\beta}_{-} \delta\Big) }{2\sqrt{j}}\Bigg]\Bigg\} \\
K &= 1 + \frac{ \varepsilon}{j} \Big[3 -  \bar{\beta}_{+} + 2 \bar{\gamma} + \bar{\beta}_{-} \delta\Big] \nn\\*
& \qquad+ \frac{\varepsilon^2}{j^2} \Bigg[\frac{105}{4} + \frac{3}{2} \bar{\beta}_{-}^2 - 15 \bar{\beta}_{+} + \frac{3}{2} \bar{\beta}_{+}^2 -  \frac{5}{4} \bar{\delta}_{+} + \frac{139}{4} \bar{\gamma} - 12 \bar{\beta}_{+} \bar{\gamma} + \frac{187}{16} \bar{\gamma}^2 - 2 \bar{\chi}_{+} \nn\\
&\qquad\qquad\quad+\delta \Big(15 \bar{\beta}_{-} - 3 \bar{\beta}_{-} \bar{\beta}_{+} -  \frac{5}{4} \bar{\delta}_{-} + 12 \bar{\beta}_{-} \bar{\gamma} + 2 \bar{\chi}_{-}\Big)  \nn\\
&\qquad\qquad\quad+ \nu\Big(- \frac{15}{2} - 6 \bar{\beta}_{-}^2 + \frac{21}{2} \bar{\beta}_{+} - 2 \bar{\delta}_{+} - 8 \bar{\gamma} -  \frac{1}{2} \bar{\gamma}^2 + 24 \bar{\beta}_{-}^2 \bar{\gamma}^{-1} - 24 \bar{\beta}_{+}^2 \bar{\gamma}^{-1} + 4 \bar{\chi}_{+} -  \frac{3}{2} \bar{\beta}_{-} \delta\Big)  \nn\\
&\qquad\qquad\quad + j\Bigg(-\frac{15}{4} +  \frac{1}{4} \bar{\delta}_{+} - \frac{15}{4} \bar{\gamma} - \frac{15}{16} \bar{\gamma}^2 +  \frac{1}{4} \bar{\delta}_{-} \delta +  \nu\Big(\frac{3}{2} -  \frac{1}{2} \bar{\beta}_{+} + \bar{\gamma} + \frac{1}{2} \bar{\beta}_{-} \delta\Big) \Bigg)\Bigg]\\
a_r&= \frac{ \tilde{G} \alpha m}{c^2 \varepsilon} \Bigg\{1 + \varepsilon \Bigg[- \frac{7}{4} -  \bar{\gamma} + \frac{1}{4} \nu\Bigg] \nn\\
&\qquad\qquad\ \ + \varepsilon^2 \Bigg[\frac{1+\nu^2}{16} +  \frac{1}{j}\Bigg(-4 + 2 \bar{\beta}_{+} +  \frac{2}{3} \bar{\delta}_{+} - \frac{16}{3} \bar{\gamma} + 2 \bar{\beta}_{+} \bar{\gamma} - \frac{11}{6} \bar{\gamma}^2 +  \frac{2}{3} \bar{\chi}_{+} +  \delta\Big(-2 \bar{\beta}_{-} + \frac{2}{3} \bar{\delta}_{-}  - 2 \bar{\beta}_{-} \bar{\gamma} -  \frac{2}{3} \bar{\chi}_{-}\Big)  \nn\\
&\qquad\qquad\qquad\qquad\qquad\qquad\quad +  \nu\Big(7 - 5 \bar{\beta}_{+}  +\frac{2}{3} \bar{\delta}_{+} + \frac{17}{3} \bar{\gamma} + \frac{1}{6} \bar{\gamma}^2 - 8 \bar{\beta}_{-}^2 \bar{\gamma}^{-1} + 8 \bar{\beta}_{+}^2 \bar{\gamma}^{-1} -  \frac{4}{3} \bar{\chi}_{+} + \bar{\beta}_{-} \delta \Big) \Bigg)\Bigg]\Bigg\} \\
e_r&= 
\sqrt{1 -  j}\Bigg\{1 + \frac{\varepsilon}{1 -  j} \Bigg[3 -  \bar{\beta}_{+} + 2 \bar{\gamma} + \bar{\beta}_{-} \delta -  \frac{1}{2} \nu + j \Bigg(- \frac{15}{8} -  \bar{\gamma} + \frac{5}{8} \nu \Bigg) \Bigg] \nn\\
&\qquad\qquad\ \ + \frac{\varepsilon^2}{(1 -  j)^2} \Bigg[- \frac{35}{4} -  \frac{1}{2} \bar{\beta}_{-}^2 + \frac{21}{4} \bar{\beta}_{+} -  \frac{1}{2} \bar{\beta}_{+}^2 + \frac{5}{2} \bar{\delta}_{+} - 12 \bar{\gamma} + 6 \bar{\beta}_{+} \bar{\gamma} -  \frac{35}{8} \bar{\gamma}^2 + 2 \bar{\chi}_{+} \nn\\
&\qquad\qquad\qquad\qquad\qquad+ \delta \Big(- \frac{21}{4} \bar{\beta}_{-} + \bar{\beta}_{-} \bar{\beta}_{+} + \frac{5}{2} \bar{\delta}_{-} - 6 \bar{\beta}_{-} \bar{\gamma} - 2 \bar{\chi}_{-}\Big)  \nn\\
&\qquad\qquad\qquad\qquad\qquad+ \nu\Big(\frac{99}{4} + 2 \bar{\beta}_{-}^2 -  \frac{65}{4} \bar{\beta}_{+} + 2 \bar{\delta}_{+} + \frac{39}{2} \bar{\gamma} + \frac{1}{2} \bar{\gamma}^2 - 24 \bar{\beta}_{-}^2 \bar{\gamma}^{-1} + 24 \bar{\beta}_{+}^2 \bar{\gamma}^{-1} - 4 \bar{\chi}_{+} + \frac{13}{4} \bar{\beta}_{-} \delta\Big)  \nn\\
&\qquad\qquad\qquad\qquad+ \frac{1}{j}\Bigg(8 - 4 \bar{\beta}_{+} -  \frac{4}{3} \bar{\delta}_{+} + \frac{32}{3} \bar{\gamma} - 4 \bar{\beta}_{+} \bar{\gamma} + \frac{11}{3} \bar{\gamma}^2 -  \frac{4}{3} \bar{\chi}_{+} + \delta\Big(4 \bar{\beta}_{-} -  \frac{4}{3} \bar{\delta}_{-} + 4 \bar{\beta}_{-} \bar{\gamma} + \frac{4}{3} \bar{\chi}_{-}\Big)  \nn\\
&\qquad\qquad\qquad\qquad\qquad+ \nu\Big(-14 + 10 \bar{\beta}_{+} -  \frac{4}{3} \bar{\delta}_{+} -  \frac{34}{3} \bar{\gamma} -  \frac{1}{3} \bar{\gamma}^2 + 16 \bar{\beta}_{-}^2 \bar{\gamma}^{-1} - 16 \bar{\beta}_{+}^2 \bar{\gamma}^{-1} + \frac{8}{3} \bar{\chi}_{+} - 2 \bar{\beta}_{-} \delta\Big)\Bigg) \nn\\
&\qquad\qquad\qquad\qquad + j \Bigg(- \frac{25}{8} -  \frac{1}{8} \bar{\beta}_{+} -  \frac{7}{6} \bar{\delta}_{+} -  \frac{41}{12} \bar{\gamma} -  \bar{\beta}_{+} \bar{\gamma} -  \frac{19}{24} \bar{\gamma}^2 -  \frac{2}{3} \bar{\chi}_{+} +  \delta\Big(\frac{1}{8} \bar{\beta}_{-} -  \frac{7}{6} \bar{\delta}_{-} + \bar{\beta}_{-} \bar{\gamma} + \frac{2}{3} \bar{\chi}_{-}\Big) \nn\\
& \qquad\qquad\qquad\qquad\qquad + \nu\Big(- \frac{37}{4} + \frac{51}{8} \bar{\beta}_{+} -  \frac{2}{3} \bar{\delta}_{+} -  \frac{89}{12} \bar{\gamma} -  \frac{1}{6} \bar{\gamma}^2 + 8 \bar{\beta}_{-}^2 \bar{\gamma}^{-1} - 8 \bar{\beta}_{+}^2 \bar{\gamma}^{-1} + \frac{4}{3} \bar{\chi}_{+} -  \frac{11}{8} \bar{\beta}_{-} \delta\Big) -  \frac{1}{16} \nu^2\Bigg) \nn\\
&\qquad\qquad\qquad\qquad + j^2 \Bigg(\frac{415}{128} + \frac{29}{8} \bar{\gamma} + \bar{\gamma}^2 + \nu\Big(- \frac{105}{64} -  \frac{7}{8} \bar{\gamma}\Big) + \frac{7}{128} \nu^2\Bigg)\Bigg]\Bigg\} \\
e_t&= \sqrt{1 -  j}\Bigg\{1 + \frac{\varepsilon}{1 -  j} \Bigg[-1 -  \bar{\beta}_{+} + \bar{\beta}_{-} \delta + \nu+ j \Bigg(\frac{17}{8} + \bar{\gamma} -  \frac{7}{8} \nu\Bigg)\Bigg] \nn\\
&\qquad \qquad\qquad \ \ + \frac{\varepsilon^2}{\sqrt{j}}\Bigg[-\frac{15}{2} +  \frac{1}{2} \bar{\delta}_{+} - \frac{15}{2} \bar{\gamma} - \frac{15}{8} \bar{\gamma}^2 +  \frac{1}{2} \bar{\delta}_{-} \delta +  \nu\Big(3 -  \bar{\beta}_{+} + 2 \bar{\gamma} + \bar{\beta}_{-} \delta\Big) \Bigg]\nn\\
&\qquad\qquad\ \ + \frac{\varepsilon^2}{(1 -  j)^2} \Bigg[- \frac{15}{4} -  \frac{1}{2} \bar{\beta}_{-}^2 + \frac{21}{4} \bar{\beta}_{+} -  \frac{1}{2} \bar{\beta}_{+}^2 + \frac{7}{6} \bar{\delta}_{+} -  \frac{41}{6} \bar{\gamma} + 4 \bar{\beta}_{+} \bar{\gamma} -  \frac{65}{24} \bar{\gamma}^2 + \frac{2}{3} \bar{\chi}_{+} \nn\\
&\qquad\qquad\qquad\qquad\qquad + \delta \Big(- \frac{21}{4} \bar{\beta}_{-} + \bar{\beta}_{-} \bar{\beta}_{+} + \frac{7}{6} \bar{\delta}_{-} - 4 \bar{\beta}_{-} \bar{\gamma} -  \frac{2}{3} \bar{\chi}_{-}\Big) + \frac{3}{4} \nu^2 \nn\\
&\qquad\qquad\qquad\qquad\qquad + \nu\Big(\frac{25}{2} + 2 \bar{\beta}_{-}^2 -  \frac{31}{4} \bar{\beta}_{+} + \frac{2}{3} \bar{\delta}_{+} + \frac{29}{3} \bar{\gamma} + \frac{1}{6} \bar{\gamma}^2 - 8 \bar{\beta}_{-}^2 \bar{\gamma}^{-1} + 8 \bar{\beta}_{+}^2 \bar{\gamma}^{-1} -  \frac{4}{3} \bar{\chi}_{+} + \frac{11}{4} \bar{\beta}_{-} \delta\Big)  \nn\\
&\qquad\qquad\qquad\qquad+ \frac{1}{j}\Bigg(4 - 2 \bar{\beta}_{+} -  \frac{2}{3} \bar{\delta}_{+} + \frac{16}{3} \bar{\gamma} - 2 \bar{\beta}_{+} \bar{\gamma} + \frac{11}{6} \bar{\gamma}^2 -  \frac{2}{3} \bar{\chi}_{+} + \delta\Big(2 \bar{\beta}_{-} -  \frac{2}{3} \bar{\delta}_{-} + 2 \bar{\beta}_{-} \bar{\gamma} + \frac{2}{3} \bar{\chi}_{-}\Big)  \nn\\
&\qquad\qquad\qquad\qquad\qquad+ \nu\Big(-7 + 5 \bar{\beta}_{+} -  \frac{2}{3} \bar{\delta}_{+} -  \frac{17}{3} \bar{\gamma} -  \frac{1}{6} \bar{\gamma}^2 + 8 \bar{\beta}_{-}^2 \bar{\gamma}^{-1} - 8 \bar{\beta}_{+}^2 \bar{\gamma}^{-1} + \frac{4}{3} \bar{\chi}_{+} -  \bar{\beta}_{-} \delta\Big)\Bigg) \nn\\
&\qquad\qquad\qquad\qquad+ j \Bigg(- \frac{45}{8} -  \frac{17}{8} \bar{\beta}_{+} -  \frac{1}{2} \bar{\delta}_{+} - 4 \bar{\gamma} -  \bar{\beta}_{+} \bar{\gamma} -  \frac{5}{8} \bar{\gamma}^2 + \delta\Big(\frac{17}{8} \bar{\beta}_{-} -  \frac{1}{2} \bar{\delta}_{-} + \bar{\beta}_{-} \bar{\gamma}\Big)  \nn\\
&\qquad\qquad\qquad\qquad\qquad +\nu\Big(- \frac{73}{16} + \frac{23}{8} \bar{\beta}_{+} -  \frac{7}{2} \bar{\gamma} -  \frac{15}{8} \bar{\beta}_{-} \delta\Big)  -  \frac{11}{8} \nu^2\Bigg) \nn\\
&\qquad\qquad\qquad\qquad+ j^2 \Bigg(\frac{607}{128} + \frac{35}{8} \bar{\gamma} + \bar{\gamma}^2 + \nu\Big(- \frac{69}{64} -  \frac{5}{8} \bar{\gamma}\Big) + \frac{79}{128} \nu^2\Bigg)\Bigg]\Bigg\} \\
e_\phi&=\sqrt{1 -  j}\Bigg\{1 + \frac{\varepsilon}{1 -  j} \Bigg[ 3 -  \bar{\beta}_{+} + 2 \bar{\gamma} + \bar{\beta}_{-} \delta + j \Bigg(- \frac{15}{8} -  \bar{\gamma} + \frac{1}{8} \nu\Bigg)\Bigg] \nn\\
&\qquad \qquad \ \ +  \frac{\varepsilon^2}{(1 -  j)^2} \Bigg[- \frac{75}{4} -  \frac{1}{2} \bar{\beta}_{-}^2 + \frac{37}{4} \bar{\beta}_{+} -  \frac{1}{2} \bar{\beta}_{+}^2 + \frac{11}{6} \bar{\delta}_{+} -  \frac{74}{3} \bar{\gamma} + 10 \bar{\beta}_{+} \bar{\gamma} -  \frac{205}{24} \bar{\gamma}^2 + \frac{10}{3} \bar{\chi}_{+} \nn\\
&\qquad\qquad\qquad\qquad\qquad+ \delta\Big(- \frac{37}{4} \bar{\beta}_{-} + \bar{\beta}_{-} \bar{\beta}_{+} + \frac{11}{6} \bar{\delta}_{-} - 10 \bar{\beta}_{-} \bar{\gamma} -  \frac{10}{3} \bar{\chi}_{-}\Big)+ \frac{33}{32} \nu^2\nn\\
&\qquad\qquad\qquad\qquad\qquad + \nu\Big(\frac{125}{32} + 2 \bar{\beta}_{-}^2 -  \frac{59}{4} \bar{\beta}_{+} + \frac{10}{3} \bar{\delta}_{+} + \frac{47}{6} \bar{\gamma} + \frac{5}{6} \bar{\gamma}^2 - 40 \bar{\beta}_{-}^2 \bar{\gamma}^{-1} + 40 \bar{\beta}_{+}^2 \bar{\gamma}^{-1} -  \frac{20}{3} \bar{\chi}_{+} + \frac{7}{4} \bar{\beta}_{-} \delta\Big) \nn\\
&\qquad\qquad\qquad\qquad  + j \Bigg(\frac{15}{8} -  \frac{17}{8} \bar{\beta}_{+} -  \frac{5}{6} \bar{\delta}_{+} + \frac{35}{12} \bar{\gamma} - 3 \bar{\beta}_{+} \bar{\gamma} + \frac{31}{24} \bar{\gamma}^2 -  \frac{4}{3} \bar{\chi}_{+} + \delta\Big(\frac{17}{8} \bar{\beta}_{-} -  \frac{5}{6} \bar{\delta}_{-} + 3 \bar{\beta}_{-} \bar{\gamma} + \frac{4}{3} \bar{\chi}_{-}\Big)  \nn\\
&\qquad\qquad\qquad\qquad\qquad  + \nu\Big(- \frac{15}{32} + \frac{47}{8} \bar{\beta}_{+} -  \frac{4}{3} \bar{\delta}_{+} -  \frac{31}{12} \bar{\gamma} -  \frac{1}{3} \bar{\gamma}^2 + 16 \bar{\beta}_{-}^2 \bar{\gamma}^{-1} - 16 \bar{\beta}_{+}^2 \bar{\gamma}^{-1} + \frac{8}{3} \bar{\chi}_{+} -  \frac{7}{8} \bar{\beta}_{-} \delta\Big)  -  \frac{21}{32} \nu^2\Bigg) \nn\\
&\qquad\qquad\qquad\qquad+ \frac{1}{j}\Bigg(13 - 6 \bar{\beta}_{+} -  \bar{\delta}_{+} + 17 \bar{\gamma} - 6 \bar{\beta}_{+} \bar{\gamma} + \frac{23}{4} \bar{\gamma}^2 - 2 \bar{\chi}_{+} + \delta\Big(6 \bar{\beta}_{-} -  \bar{\delta}_{-} + 6 \bar{\beta}_{-} \bar{\gamma} + 2 \bar{\chi}_{-}\Big)\nn\\
&\qquad\qquad\qquad\qquad\qquad + \nu\Big(- \frac{91}{32} + 9 \bar{\beta}_{+} - 2 \bar{\delta}_{+} - 5 \bar{\gamma} -  \frac{1}{2} \bar{\gamma}^2 + 24 \bar{\beta}_{-}^2 \bar{\gamma}^{-1} - 24 \bar{\beta}_{+}^2 \bar{\gamma}^{-1} + 4 \bar{\chi}_{+} -  \bar{\beta}_{-} \delta\Big) -  \frac{15}{32} \nu^2\Bigg)\nn\\
&\qquad\qquad\qquad\qquad + j^2 \Bigg(\frac{415}{128} + \frac{29}{8} \bar{\gamma} + \bar{\gamma}^2 + \nu\Big(- \frac{47}{64} -  \frac{3}{8} \bar{\gamma}\Big)  + \frac{11}{128} \nu^2\Bigg)\Bigg]\Bigg\}\\
f_t&= \frac{\sqrt{1 -  j} \varepsilon^2 \nu (15 + 8 \bar{\gamma} - \nu)}{8 \sqrt{j}}\\
f_\phi&=\frac{(1 -  j) \varepsilon^2 \Big(4 + 4 \bar{\delta}_{+} + 4 \bar{\gamma} + \bar{\gamma}^2 + 4 \bar{\delta}_{-} \delta +\big(76 - 24 \bar{\beta}_{+} + 48 \bar{\gamma} + 8 \bar{\beta}_{-} \delta \big) \nu - 12 \nu^2\Big)}{32 j^2}\\
g_t&=\frac{\varepsilon^2 \Big(60 - 4 \bar{\delta}_{+} + 60 \bar{\gamma} + 15 \bar{\gamma}^2 - 4 \bar{\delta}_{-} \delta + \nu \big(-24 + 8 \bar{\beta}_{+} - 16 \bar{\gamma} - 8 \bar{\beta}_{-} \delta\big) \Big)}{8 \sqrt{j}} \\
g_\phi&=\frac{ \varepsilon^2  (1 -  j)^{3/2} \nu (1 - 3 \nu)}{32 j^2}
\end{align}\ese
\end{widetext}

\bibliography{references.bib}

\end{document}